\documentclass[a4paper,11pt]{article}
\pdfoutput=1 

\usepackage{jcappub} 

\usepackage[T1]{fontenc} 
\usepackage{xcolor}
\usepackage{graphicx} 
\usepackage{float}
\usepackage{multirow}

\usepackage{arydshln}
\usepackage{booktabs}

\newcommand{\be}{\begin{equation}}
\newcommand{\ee}{\end{equation}}
\newcommand{\bq}{\begin{eqnarray}}
\newcommand{\eq}{\end{eqnarray}}
\newcommand{\vv}[0]{{\bar v}}
\newcommand{\cc}[0]{{\tilde c}}
\newcommand{\xx}[0]{{\mathbf{x}}}
\newcommand{\xd}[0]{{\dot{\mathbf{x}}}}
\newcommand{\xp}[0]{{\mathbf{x}}'}
\newcommand{\xhp}[2]{\hat{{\rm x}}'^{(#2)}_{#1}}
\newcommand{\xhd}[1]{\hat{\dot{{\rm x}}}_{#1}}
\newcommand{\un}[1]{\hat{\mathbf{#1}}}
\newcommand{\uni}[2]{\hat{{{\rm #1}}}_{#2}}


\title{
\begin{center} 
Cosmic strings and domain walls: 
the impact of CMB $B$-mode data
\end{center}
}

\author[a]{Luca~Caloni,}
\author[a,b]{Ricardo~Z.~Ferreira,}
\author[d,c]{Lara~Sousa,}
\author[c,d]{and Clara~Winckler}

\affiliation[a]{Faculdade de Ci\^{e}ncias e Tecnologia and CFisUC, Departamento de F\'isica,\\ 
Universidade de Coimbra, Rua Larga P-3004-516 Coimbra, Portugal}
\affiliation[b]{Centro de F\'isica das Universidades do Minho e do Porto (CF-UM-UP),\\
Universidade do Minho, P-4710-057 Braga, Portugal}
\affiliation[c]{Departamento de Física e Astronomia, Faculdade de Ciências,\\
Universidade do Porto, Rua do Campo Alegre 687, PT4169-007 Porto, Portugal}
\affiliation[d]{Instituto de Astrofísica e Ciências do Espaço,\\ Universidade do Porto,
CAUP, Rua das Estrelas, PT4150-762 Porto, Portugal}

\emailAdd{luca.caloni@uc.pt}
\emailAdd{rzferreira@fisica.uminho.pt}
\emailAdd{lara.sousa@astro.up.pt}
\emailAdd{clara.winckler@astro.up.pt}

\abstract{We analyse CMB constraints on stable networks of cosmic strings and domain walls using for the first time full {\it Planck} 2018 data together with BICEP/{\it Keck} 2018 $B$-mode measurements. The defect-induced anisotropies are computed using the Unconnected Segment Model for Nambu-Goto and Abelian-Higgs strings, as well as for stable domain walls, and included in a full Markov Chain Monte Carlo analysis jointly varying all $\Lambda$CDM parameters, the tensor-to-scalar ratio, and the string/domain wall tension. No statistically significant evidence for defects is found, although we observe a mild preference for non-zero cosmic string tension. Our results improve previous constraints on the cosmic string power spectrum and correct previous CMB analyses with domain walls that overestimated the constraints. In the particular case of strings, the improvement is driven by the $B$-mode data, and is especially pronounced for Abelian-Higgs strings. We also present forecasts for the Simons Observatory and the \textit{LiteBIRD} satellite: the former will improve string tension constraints by about a factor of three, while the latter those on the domain wall tension by about a factor of five. 
Finally, we assess the impact of Nambu-Goto string loops on CMB anisotropies in light of both current and future observations.}

\begin{document}
\maketitle
\flushbottom


\section{Introduction}

In the search for new physics, one is often tempted to probe smaller and smaller length scales. There are, however, scenarios in which new physics manifests itself in macroscopic structures of cosmological size. Networks of cosmic strings and domain walls --- features that arise naturally in many well-motivated extensions of the Standard Model of particle physics (see e.g.~\cite{Vilenkin:2000jqa,Vachaspati:2006zz}) --- are two such examples.

One way to probe such cosmic defects is to study their impact on the Cosmic Microwave Background (CMB). Most searches for cosmic defects in the CMB have been focused on cosmic strings, dating back to when they were still considered a possible source of the primordial fluctuations observed on the largest scales of our Universe. Although this possibility was ruled out by WMAP observations~\cite{WMAP:2003elm}, the motivation to search for cosmic strings has remained intact (see e.g.~\cite{Albrecht:1997mz,Pogosian:2003mz,Battye:2006pk,Seljak:2006hi,Hindmarsh:2009qk,Regan:2010cn,Lazanu:2014xxa}).
The state of the art in this respect goes back to the {\it Planck} 2015 analysis of temperature and polarization data~\cite{Planck:2015fie}, later refined in~\cite{Lizarraga:2016onn,Charnock:2016nzm,Rybak:2021scp}. 
No evidence for cosmic strings has been found, leading instead to upper bounds on the string tension $\mu$, normalized by Newton’s constant, at the level of $(1$–$2)\times10^{-7}$, depending on the specific string model considered. 
In contrast, CMB searches for domain wall networks are much rarer and, to our knowledge, have only been done in~\cite{Lazanu:2015fua,Sousa:2015cqa}, yielding an upper bound on the domain wall tension $\sigma$ of around $(0.9\,\rm{MeV})^3$.   

In this work, we revisit and update the CMB analysis of cosmic strings and domain walls, incorporating for the first time \textit{Planck} 2018 and BICEP/\textit{Keck} 2018 data~\cite{Planck:2018nkj,BICEP:2021xfz}, together with CMB lensing measurements from \textit{Planck} PR4~\cite{Carron:2022eyg}. As we will show, the new data, in particular the inclusion of CMB $B$-mode polarization, provides significant additional constraining power.

We will focus on the gravitational effects of the defects on the CMB, encoded in the unequal-time correlators (UETC) of their stress-energy tensor that act as active sources of perturbations (see also~\cite{Agrawal:2019lkr,Pogosian:2019jbt,Jain:2021shf,Bortolami:2022whx,Jain:2022jrp,Ferreira:2023jbu}, for some recent studies exploring non-gravitational probes). Two main approaches have been employed in the literature to determine these correlators: (i) numerically, via large field theory or Nambu-Goto simulations\footnote{The restricted dynamical range of the simulations often requires extrapolation to a substantially longer temporal range.}~\cite{Lazanu:2014xxa,Lazanu:2015fua,Lizarraga:2016onn,Lopez-Eiguren:2017dmc}; and (ii) semi-analytically, using the so-called unconnected segment model (USM), in which cosmic string and domain wall networks are modeled as a collection of randomly oriented infinitely-thin flat segments/sections with scaling properties described by the VOS model~\cite{Martins:1996jp,Martins:2000cs,Avelino:2005kn,Avelino:2011ev,Sousa:2011ew,Sousa:2011iu} (that is calibrated to match the large-scale dynamics observed in numerical simulations~\cite{Albrecht:1997mz,Pogosian:1999np,Lizarraga:2016onn,Sousa:2015cqa,Charnock:2016nzm}).

Here, we will follow this semi-analytical approach and use the USM to compute the stress-energy tensor of the strings and walls which we then supplement to the \texttt{CMBACT4} code~\cite{Pogosian:1999np,CMBACT} to compute the CMB spectra generated by these active sources. 
In the case of cosmic strings, we will consider separately the case of Nambu-Goto and Abelian-Higgs strings, due to the known discrepancies between the results of the two different types simulations ~\cite{Hindmarsh:2017qff,Blanco-Pillado:2023sap}. These are in part justified by the fact that the field theory (Abelian-Higgs) simulations include additional radiative channels, thus also serving, to a first approximation, as a proxy for other radiative models such as global strings~\cite{Planck:2013mgr} or superconducting strings~\cite{Rybak:2024djq}. We also include for the first time the effects of string loops on the full data analysis following the methodology derived in~\cite{Rybak:2021scp}.

After generating the defect spectra, we perform a full Markov Chain Monte Carlo (MCMC) analysis by sampling over the standard $\Lambda$CDM parameters, the tensor-to-scalar ratio $r$, and the cosmic string or domain wall tension. 
Finally, we present the first dedicated forecasts for cosmic defect searches with upcoming CMB surveys, including the Simons Observatory~\cite{SimonsObservatory:2018koc} and the \textit{LiteBIRD}~\cite{LiteBIRD:2022cnt} satellite, and discuss the relevance of current and future CMB data for probing cosmic defects.

The paper is organized as follows.
In section~\ref{sec:USM} we describe how we model the evolution of cosmic string and domain wall networks based on the USM and VOS models. 
The impact of these defect networks on CMB anisotropies is then discussed in section~\ref{sec:CMB-anisotropies}.
In section~\ref{sec:constraints} we present the constraints obtained from a full MCMC analysis using \textit{Planck} 2018 and BICEP/\textit{Keck} 2018 data, while section~\ref{sec:forecasts} presents forecasts for future CMB observations with the Simons Observatory and the {\it LiteBIRD} satellite. In section \ref{sec:Implications for other non-CMB GW observations} we discuss the implications of the results for other searches for defects in gravitational wave observatories.
Finally, we summarize our findings and discuss future prospects in section~\ref{sec:conclusions}.

\section{Modelling Cosmic Defects with the Unconnected Segment Model}
\label{sec:USM}

Topological defects, unlike inflation, source perturbations actively from their production in the early universe until the present time (or their time of decay, if the networks are meta-stable). Computing the CMB anisotropies sourced by defects then requires a characterization of the stress-energy tensor of the whole network from the pre-recombination era to the present times. 
This may be achieved either by extrapolating the results of numerical simulations or by resorting to an analytical framework known as the Unconnected Segment Model (USM) --- first introduced for cosmic strings in~\cite{Albrecht:1997mz,Pogosian:1999np} and later extended for domain walls in~\cite{Sousa:2015cqa}. Here, we will follow the latter approach and use the publicly available \texttt{CMBACT4} code~\cite{Pogosian:1999np} to compute the temperature and polarization anisotropies. We will consider three distinct defect scenarios: stable domain walls, Nambu-Goto strings (with and without string loops) and Abelian-Higgs strings. In this section, we describe how each of these scenarios is implemented in \texttt{CMBACT4}. We start with a review of the semi-analytical models describing the cosmological evolution of each type of defect.
We then outline the essential features of the USM for cosmic string and domain wall networks.

\subsection{Cosmological evolution of topological defect networks}
The formation of topological defects in the early universe is a generic prediction of particle physics scenarios in which there is spontaneous symmetry breaking~\cite{Kibble:1976sj,Vilenkin:2000jqa,Vachaspati:2006zz}. They appear in regions in which the fields cannot, by continuity, relax into a vacuum and are thus non-trivial field configurations with an associated energy density. The dimensionality of the defects depends on the type of symmetry that is broken (or, more precisely, on the topology of the vacuum manifold of the theory). Sheet-like defects, known as domain walls, may form whenever there is a breaking of a discrete symmetry; line-like defects, dubbed cosmic strings, may be produced if an axial symmetry is broken; and monopoles (point-like defects) may arise in the breaking of a spherical symmetry. Here, we focus our analysis on domain walls and cosmic strings.

 An infinitely-thin $p$-dimensional object traces out a $p$$+1$-dimensional worldvolume in spacetime:
\begin{eqnarray}
X^\nu=X^\nu (\zeta^a)\,,\quad\mbox{with}\quad a=0,1,\cdots,p\,,
\end{eqnarray}
where $\zeta^0$ is a timelike coordinate and $\zeta^1,\cdots,\zeta^p$ are spacelike coordinates that parameterize the worldvolume. Their dynamics may then be described by the generalized Nambu-Goto (NG) action~\cite{Vilenkin:2000jqa,Sousa:2011iu}:
\begin{eqnarray}
S=-\sigma_p\int\sqrt{-\gamma}d^{p+1}\zeta\,,
\label{eq:NGaction}
\end{eqnarray}
where $\sigma_p$ is the defect energy per unit $p$-dimensional area, $\gamma=\det(\gamma_{ab})$, $\gamma_{ab}=g_{\alpha\nu} X^\alpha_{,a} X^\nu_{,b}$ and $g_{\alpha\nu}$ are, respectively, the induced and the background metrics and ${}_{,a}\equiv \partial/\partial\zeta^a$.\footnote{Note that the energy per unit $p$-dimensional area of the defect is determined by the Lagrangian density $\mathcal{L}$ of the underlying field theory model~\cite{Vilenkin:2000jqa,Sousa:2011iu}:
\begin{eqnarray}
\sigma_p=-\int d^{3-p}x \mathcal{L}\,.
\end{eqnarray}
In the infinitely-thin limit, the dynamics of the defects do not depend on $\mathcal{L}$.}

To describe the cosmological evolution of a defect network of dimension $p$, we resort to the Velocity-dependent One-Scale (VOS) model, which was first introduced to describe cosmic string networks~\cite{Martins:1996jp,Martins:2000cs} and later extended to domain walls~\cite{Avelino:2005kn,Avelino:2011ev}, monopoles~\cite{Martins:2008zz,Sousa:2017wvx} and to defects of arbitrary dimensionality~\cite{Sousa:2011ew,Sousa:2011iu}. This model provides a thermodynamical description of the evolution of the network in terms of two variables: the characteristic lengthscale of the network $L$ --- defined as
\be
\rho=\frac{\sigma_p}{L^{3-p}}\,,
\ee
where $\rho$ is the average energy density of the network --- and the Root-Mean-Squared (RMS) velocity $\vv$. This characteristic length is not only a measure of $\rho$, but also roughly gives us the typical distance between defects, their correlation length and their typical curvature radii. For a network of defects of dimensionality $p$, the evolution equations for $\vv$ and $L$ take the form~\cite{Sousa:2011iu}
\bq
\dot{\vv} & = & (1-\vv^2)\left[\frac{k(\vv)}{L_c}-(p+1)\frac{\dot{a}}{a}\vv \right]\,,\label{eq:VOSv}\\
\dot{L}_c & = & \frac{p+1}{3-p}\frac{\dot{a}}{a}L_c\vv^2+ \frac{\cc\vv}{3-p}\,,\label{eq:VOSL}
\eq
where $L_c=L/a$ is the characteristic conformal length and 
dots represent derivatives with respect to conformal time $\tau$. These equations may be derived by averaging the generalized Nambu-Goto equations of motion (obtained by varying~\eqref{eq:NGaction} with respect to $X^\nu$) over the whole network. Note however that $k(\vv)$ is a phenomenological curvature/momentum parameter, describing the average acceleration felt by the defects, and the second term in~\eqref{eq:VOSL} is a phenomenological term added to describe the energy loss in the network 
(which are not included in~\eqref{eq:NGaction}).

We now proceed with concrete cases of cosmic strings and domain walls.

\subsubsection{Cosmic strings}

As anticipated at the beginning of this section, we consider two different models for cosmic strings: Nambu-Goto (NG) strings and Abelian-Higgs (AH) strings. The Abelian-Higgs (AH) model is described by a Lagrangian density
\begin{eqnarray}
\mathcal{L}=\left|D_\alpha \phi\right|^2-\frac{1}{4e^2}F_{\alpha\nu}F^{\alpha\nu}-\frac{\lambda}{4}\left(\left|\phi\right|-\eta^2\right)^2\,,
\label{eq:AHmodel}
\end{eqnarray}
where $\phi$ is complex scalar field, $e$ and $\lambda$ are dimensionless coupling constants, $D_\mu=\partial_\alpha+iA_\alpha$, $F_{\alpha\nu}=\partial_\alpha A_\nu-\partial_\nu A_\alpha$ is an anti-symmetric tensor and $A_\alpha$ is the gauge vector field. This model has a local (gauge) U(1) symmetry and admits string solutions with an energy per unit length, or tension, $\mu\sim\eta^2$ that is concentrated within a thin core of width $\delta\sim 1/\eta$.

In a cosmological setting, when considering lengthscales that are much larger than $\delta$, these AH strings may be approximately treated as infinitely-thin 1-dimensional objects. The dynamics of infinitely-thin cosmic strings, or NG strings, may be described by the action in~\eqref{eq:NGaction} with $p=1$ and $\mu\equiv\sigma_1$. Under this approximation, the numerical and analytical study of cosmic string networks becomes more tractable, as the number of degrees of freedom in the problem is reduced. This has allowed, for instance, for much larger simulations of cosmic string network dynamics~\cite{Blanco-Pillado:2011egf,CamargoNevesdaCunha:2022mvg}, as  field theory simulations~\cite{Hindmarsh:2017qff,Correia:2020yqg} are  computationally more demanding. 

Although NG strings are often regarded as an effective model for AH strings (or other strings in which energy is concentrated within a thin core), this is still a matter of active debate in the literature. A common picture regarding the large-scale dynamics of cosmic strings has emerged from both NG and AH numerical simulations: string networks quickly evolve towards a linear scaling regime in which the cosmic string density $\rho$ remains a fixed fraction of the background energy density (i.e., the characteristic lengthscale grows linearly with time) and the RMS velocity remains constant\footnote{Note that such a regime is an attractor of the VOS equations~\eqref{eq:VOSv}-\eqref{eq:VOSL} and, therefore, it is also naturally included in our semi-analytical approach.} (albeit, as may be seen in figure~\ref{fig:VOSEvolution}, AH networks are predicted to be less dense and move with a smaller RMS velocity; see also~\cite{Hindmarsh:2017qff}). NG and AH simulations, however, disagree on the nature of the energy loss mechanism that enables the attainment of this linear scaling regime. In NG simulations, scaling is reached through the production of closed loops of string~\cite{Lorenz:2010sm,Blanco-Pillado:2011egf,Blanco-Pillado:2013qja}, which are expected to slowly decay by emitting gravitational radiation.\footnote{Note that current NG simulations only include gravity at the background level and do not model this decay. The emission of gravitational waves by NG string loops and the impact of gravitational backreaction have however been extensively studied analytically and numerically in the literature (see e.g.~\cite{Garfinkle:1988yi,Burden:1985md,Damour:2001bk,Wachter:2024aos}).} In the NG approximation, however, strings are assumed to have no internal degrees of freedom --- which should be a good approximation wherever/whenever string thickness is much smaller than the string's curvature radius --- while AH-like strings may also decay by emitting scalar and gauge radiation. Although it has been argued that this mechanism should play a negligible role in the evolution of the network on cosmological scales~\cite{Blanco-Pillado:2023sap}, current AH simulations indicate that it is
the direct emission of scalar and gauge radiation that ensures the attainment of the linear scaling regime~\cite{Hindmarsh:2017qff}. AH numerical simulations also show that the loops that naturally form in the evolution are typically short-lived, as they have high-curvature regions in which the emission of massive radiation occurs~\cite{Hindmarsh:2021mnl,Blanco-Pillado:2023sap,Baeza-Ballesteros:2024sny} (although they are well described by the NG action outside of these regions~\cite{Blanco-Pillado:2023sap}). Since the potential role of massive radiation in realistic cosmic string networks in a cosmological setting is still a matter of debate, here we will consider these two string scenarios separately as is often done in the literature.

\paragraph{Nambu-Goto string networks:}
 Current numerical simulations of NG string networks are well described by the VOS model in eqs.~\eqref{eq:VOSv}-\eqref{eq:VOSL}, with $p=1$ and $\cc=0.23$ and by the following \textit{ansatz} for $k(\vv)$~\cite{Martins:2000cs}:\footnote{In the original \texttt{CMBACT4}~\cite{Pogosian:1999np}, a simpler form was chosen for $k(\vv)$, along with different values of $\cc$ in the radiation and matter eras (with a fast interpolation between the them), based on an earlier calibration of the VOS model. Here, we will use the calibration in~\cite{Martins:2000cs} (that was published after~\cite{Pogosian:1999np}) as it provides a good description for both radiation and matter era simulations.}
\begin{eqnarray}
k(\vv)=\frac{2\sqrt{2}}{\pi}\frac{1-8\vv^6}{1+8\vv^6} \left(1-\vv^2\right)\left(1+2\sqrt{2}\vv^3 \right)\,,
\end{eqnarray}
which was constructed based on analytical results obtained for a helicoidal string solution in the relativistic and non-relativistic regimes. In this scenario, $\cc$ may be interpreted as a parameter that characterizes the efficiency of loop-chopping in string collisions and self-interactions. NG simulations show that about $10\%$ of the energy lost as a result of this loop production is in the form of large long-lived loops with conformal lengths $l_c\simeq 0.34 L_c$ and translational velocities of $v_l\sim 0.42$~\cite{Lorenz:2010sm,Blanco-Pillado:2013qja}. The rest of the energy loss is in the form of small short-lived loops with high velocities that should have a negligible impact on the CMB anisotropies~\cite{Rybak:2021scp}. In figure~\ref{fig:VOSEvolution}, we show the evolution of $L$ and $\vv$ predicted by this VOS model for NG strings in a $\Lambda$CDM background (solid blue line). Note that, because the VOS equations are solved alongside the Friedmann equation, the radiation-matter and matter-$\Lambda$ transitions arise naturally in this model. In particular, we see that the radiation-matter transition is quite slow and the network does not have enough time to re-establish a linear scaling regime before $\Lambda$ becomes dynamically important.

\paragraph {Abelian-Higgs string networks:} Although the VOS model was derived within the NG approximation, additional phenomenological terms may be added to account for the microphysics of the AH model. Here, we describe the evolution of a network of AH strings using one such model, the extended VOS model introduced in~\cite{Correia:2019bdl}. The evolution equations, in this case, are also given by eqs.~\eqref{eq:VOSv} and~\eqref{eq:VOSL} with $p=1$, but now $k(\vv)$ is parameterized as
\begin{equation}
    k(\vv)=k_0\frac{1-\left(q\vv^2\right)^\beta}{1+\left(q\vv^2\right)^\beta} \, ,
    \label{eq:k-AH}
\end{equation}
where $k_0$, $q$ and $\beta$ are free parameters. Moreover, in addition to the loss of energy due to the production of string loops (the second term in eq.~\eqref{eq:VOSL}), which takes the same form as in the NG case, a term accounting for the emission of non-gravitational radiation of the form
\begin{equation}
    \left.\dot{L}_c\right|_{\rm radiation}=\frac{1}{2}d\left[k_{0}-k(\vv)\right]^s \, ,\label{eq:vosl-AH}
\end{equation}
is added to~\eqref{eq:VOSL}. Here, $d$ and $s$ are additional free parameters characterizing the energy loss efficiency. This model therefore has a total of six free parameters that need to be calibrated using simulations. We use the calibration based on simulations with a lattice size of $4096^3$ and a step size of $\Delta x=0.25$ derived by~\cite{Correia:2021tok} and reproduced in table~\ref{tab:AHcalib}. We show the evolution of an AH string network in~figure~\ref{fig:VOSEvolution} (solid orange line). We can see that, as anticipated, $L$ is larger and $\vv$ is smaller than what is expected for NG strings. The radiation-matter transition is also slow in this case, but the network is very close to reaching the matter-era scaling regime before the onset of the $\Lambda$-dominated era.
\begin{table}[!t]
    \centering
    \resizebox{0.5\textwidth}{!}{
    \begin{tabular}{ccccccc}
        \hline      
        \hline 
        \noalign{\vskip 1mm}
        Parameter & $d$ & $s$ & $\beta$ & $k_0$ & $q$ & $\tilde{c}$\\
        \noalign{\vskip 1mm}
        \hline
        \noalign{\vskip 1mm}
        & $0.36$ & $2.56$ & $2.69$ & $1.04$ & $2.47$ & $0.30$\\[1mm]
        \hline
        \hline 
    \end{tabular}}
    \caption{Calibration of the extended VOS model as given in~\cite{Correia:2021tok}, for a lattice size of $4096^3$ and a step size of $\Delta x=0.25$.}
    \label{tab:AHcalib}
\end{table}

\subsubsection{Domain walls}
In the case of domain walls, a prototypical model that admits domain wall solutions is the Goldstone model, described by the Lagrangian
\be
\mathcal{L}=\frac{1}{2}\partial_\nu\phi\partial^\nu\phi+\frac{\lambda}{4}\left(\phi^2-\eta^2\right)^2\,,
\ee
where $\lambda$ again is a dimensionless coupling constant. Domain Walls (DWs) correspond to thin surfaces ---  with a thickness $\delta\sim1/\eta$ --- that divide space into domains with different vacuum expectation values. They are characterized by a surface tension, or energy per unit area, such that $\sigma\sim\eta^3$.

As in the case of cosmic strings, in a cosmological setting wall thickness is typically much smaller than its curvature radius, allowing the walls to be treated as infinitely-thin surfaces. In this case, walls sweep out a 2$+$1-dimensional worldsheet in spacetime whose dynamics may also be described by eq.~\eqref{eq:NGaction} with $p=2$ and $\sigma\equiv \sigma_2$. 
Numerical studies of DW network evolution currently rely on field theory simulations, which show that the network evolves towards a linear scaling regime in which the characteristic length grows linearly with time and the RMS velocity remains constant (see e.g.~\cite{Correia:2017aqf,Ferreira:2024eru,Notari:2025kqq,Blasi:2025tmn,Dankovsky:2025pjg} for recent work). However, contrary to strings, the energy density of domain walls grows linearly with time relative to the background: $\Omega_{\rm DW} \equiv \rho_{\rm DW}/\rho_c \propto t$. 

\begin{figure}
    \centering
    \includegraphics[width=0.49\linewidth]{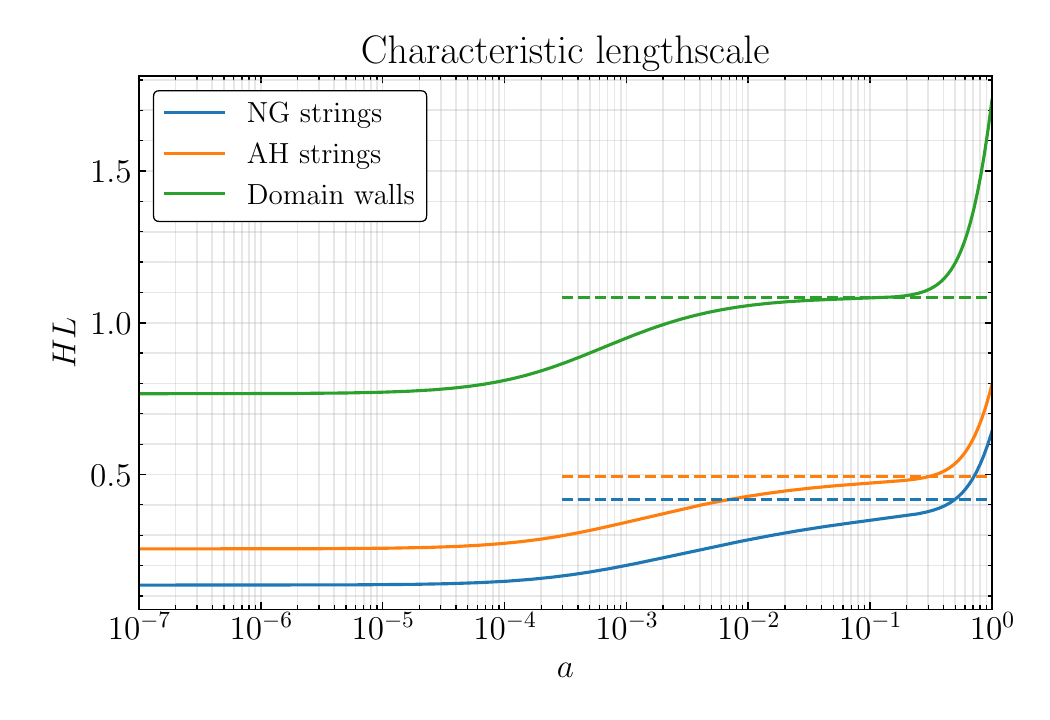}
    \includegraphics[width=0.49\linewidth]{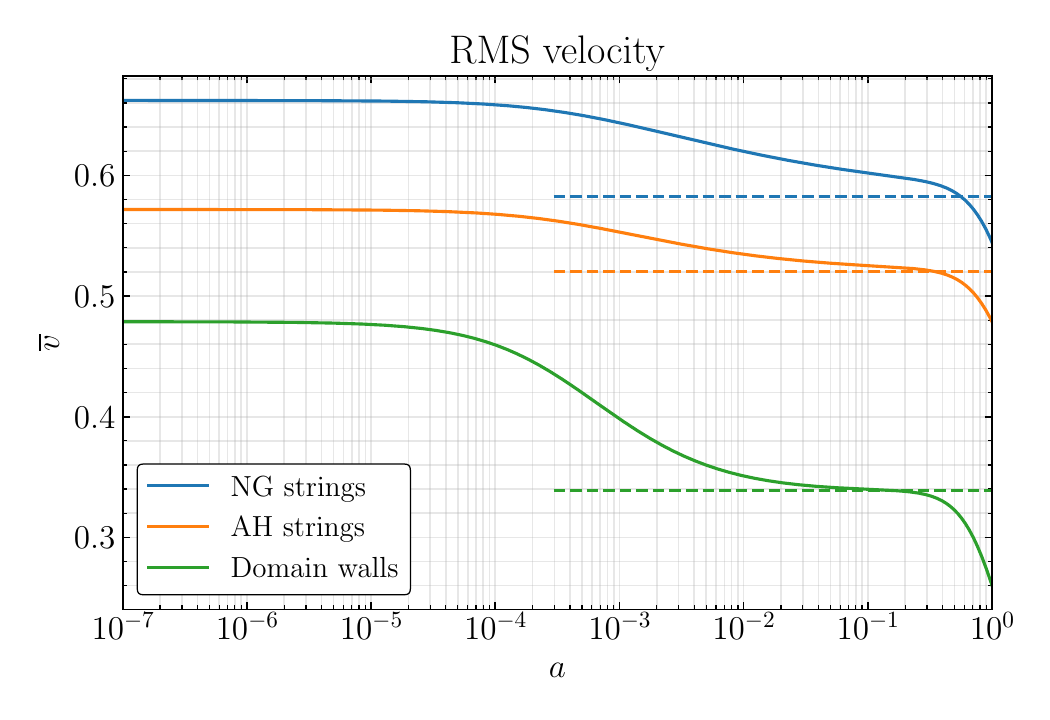}
    \caption{Evolution of the characteristic lengthscale multiplied by the Hubble parameter $LH$ (left panel) and the RMS velocity $\vv$ (right panel) as predicted by the VOS models for Nambu-Goto strings (solid blue), Abelian-Higgs strings (solid orange) and domain walls (solid green) in a $\Lambda$CDM background. The dashed lines correspond to the matter-era scaling solutions for each model.}
    \label{fig:VOSEvolution}
\end{figure}

Although there are no NG domain wall simulations,\footnote{There are NG simulations of domain wall networks but only in 2$+$1-dimensions~\cite{Blanco-Pillado:2022rad}.} the NG description is still very useful to perform analytical studies, in particular, as we shall see, to model the stress-energy tensor of the network. 
The cosmological evolution of DW networks may also be described using the VOS model in eqs.~\eqref{eq:VOSv} and~\eqref{eq:VOSL} with $p=2$ (see also~\cite{Sousa:2010zza,Avelino:2011ev} for the derivation directly from the equations of motion of the scalar field).\footnote{In the NG approximation, the general form of the equations of motion does not depend on the specific potential chosen in this thin-wall limit. However, one expects corrections to arise when the thickness is comparable to the wall's curvature radii and the microphysics becomes important~\cite{Press:1989yh,Arodz:1993sy}. In fact, most recent field theory simulations show appreciable levels of scalar radiation being produced from the network thus motivating a dedicated study of the issue~\cite{Ferreira:2024eru}.} 
Here, we will use the calibration of the VOS equations proposed in~\cite{Leite:2011sc}, that found that $\cc=0.5$ and $k(\vv)\equiv k=1.1$ provide a good fit for the evolution during the radiation-, matter- and dark-energy-dominated eras.\footnote{A more complex model, including additional radiative terms as in the case of AH strings, was developed in~\cite{Martins:2016ois} with the inclusion of several additional free parameters. 
We have verified that this model leads to larger CMB anisotropies, especially at small scales.
Here, we opted for the simpler model proposed in~\cite{Leite:2011sc} as it should allow us to derive more conservative (and safer) constraints.} The numerical solution to the VOS equations for domain walls in a $\Lambda$CDM background is shown in figure~\ref{fig:VOSEvolution} (solid green line). We find that $L$ is typically larger for DWs than for strings, implying a smaller number density of walls, corresponding roughly to one per Hubble volume. Moreover, the radiation-matter transition is faster for DWs and the network seems to be able to re-establish scaling in the matter era (albeit for a very short period of time).

\subsection{The Unconnected Segment/Section Model}
\label{subsec:USM}
In this phenomenological framework, the stress-energy tensor of cosmic defects is constructed by modelling the network as a collection of unconnected straight segments, in the case of strings, or flat square sections in the case of walls~\cite{Albrecht:1995bg,Albrecht:1997mz,Pogosian:1999np}.
All string (wall) segments (sections) are assumed to have the same comoving length (area) $L_c$ ($L_c^2$), determined by the corresponding VOS model to ensure an accurate representation of evolution of the energy density of the network.
The segments or sections are uniformly distributed in space, with randomly oriented velocities of equal magnitude given by the RMS velocity from the corresponding VOS model.
The CMB anisotropies are then computed with \texttt{CMBACT4} by performing an ensemble average over many such random realizations of the network. The results obtained using this framework have been shown to be robust and, in fact, its predictions for the angular power spectra are in good agreement with those obtained using NG~\cite{Planck:2013mgr,Lazanu:2014eya,Charnock:2016nzm}, AH~\cite{Planck:2013mgr,Correia:2021tok} and DW~\cite{Lazanu:2015fua,Sousa:2015cqa} simulations.

To account for the energy loss experienced by the network as it evolves --- caused by loop production in NG string networks, emission of scalar and gauge radiation in AH networks and by the emission of scalar radiation/collapse of sub-horizon domains in wall networks --- a fraction of the segments/sections decays at each time step $\tau_i$:
\begin{equation}
    \mathcal{N}(\tau_i) = V \left[ n(\tau_{i-1}) - n(\tau_i) \right] \, ,
    \label{eq:Ni}
\end{equation}
where $\mathcal{N}(\tau_i)$ is the number of segments/sections decaying between $\tau_{i-1}$ and $\tau_i$, $V$ is the simulation volume and $n(\tau)$ gives the number density of defects at a conformal time $\tau$. 
To ensure that this number density follows the predictions of numerical simulations, the corresponding VOS model is used to set the evolution of $n(\tau)$. In particular, one takes
\begin{eqnarray}
n(\tau)=\frac{C(\tau)}{L_c(\tau)^3}\,,
\end{eqnarray}
where the normalization $C(\tau)$ is determined by requiring that $n(\tau)$ is well described by $V/L_c(\tau_i)^3$ at each time step.

The final outcome of the USM model is the total stress-energy tensor of the defect network, which may be written as\footnote{In~\cite{Pogosian:1999np}, this expression includes an additional term that accounts for the contribution of the defects that have not decayed by the present time $\tau_0$. In \texttt{CMBACT4}, however, the simulation runs until a time $\tau>\tau_0$ and this term is ``absorbed'' into this summation.}
\begin{equation}
    \label{eq:USM}
    \Theta_{\mu\nu}^{\rm network}(\mathbf{k},\tau) = \sum_{i=1}^{N}\sqrt{\mathcal{N}(\tau_i)} \Theta_{\mu\nu}^i(\mathbf{k},\tau) T^{\rm off}(\tau,\tau_i,L_f) \, ,
\end{equation}
where the function 
\begin{align}
\label{eq:Toff}
    T^{\rm off}(\tau,\tau_i,L_f) = 
        \begin{cases}
            1 &\tau < L_f \tau_i \\
            \frac{1}{2} + \frac{1}{4}(y^3 - 3y) \quad &L_f \tau_i \le \tau < \tau_i \\
            0 & \tau_i \le \tau
        \end{cases} \, ,
\end{align}
with 
\begin{equation}
    y = 2 \frac{\ln(L_f\tau_i/\tau)}{\ln(L_f)} - 1 \, ,
\end{equation}
``turns off'' the contribution of the segments/sections that decay at a time step $\tau_i$ whenever $\tau>\tau_i$. The parameter $L_f \le 1$ controls the speed at which these segments decay and we discuss the impact of varying $L_f$ in appendix~\ref{appendix:triangle_plots}. Note that in eq.~\eqref{eq:USM}, we did not include the effect of string loops. We describe their respective stress-energy tensor and set out how these will be included in NG string networks in \texttt{CMBACT4} in appendix~\ref{app:loops} (since the loops produced in AH simulations are short-lived and thus should contribute negligibly to the anisotropies, we will not include loops in that case).

For computational efficiency, all segments/sections that decay at each timestep are described by a single (\textit{consolidated}) segment and contribute to the stress-energy tensor with a weight of $\mathcal{N}(\tau_i)^{1/2}$. This is justified by the fact that segments/sections that decay at the same time are randomly distributed in real space (i.e., they have random phases in Fourier space) and, as result, the summation of their contributions may be regarded as a random walk in two dimensions with $\mathcal{N}(\tau_i)$ steps. The summation in eq.~\eqref{eq:USM} is then performed over \textit{consolidated} segments/sections, with $N$ being the number of time steps considered in the simulations, and $\Theta_{\mu\nu}^i(\mathbf{k},\tau)$ the stress-energy tensor of the consolidated segment/section that decays at $\tau_i$. 

With this setup, the only ingredient missing to model the stress-energy tensor of the whole defect network is $\Theta_{\mu\nu}^i(\mathbf{k},\tau)$. 
We then resort to the NG action in eq.~\eqref{eq:NGaction} and, by varying it with respect to the metric tensor, one finds the energy-momentum tensor of an infinitely-thin $p$-dimensional topological defect to be given by:
\begin{equation}
    T^{\mu\nu} = \frac{\sigma_p}{\sqrt{-g}} \int d^{p+1}\zeta \left( \sqrt{-\gamma} \gamma^{ab} X^\mu_{,a} X^\nu_{,b}\right) \delta^{(4)}\left[X^\mu - X^\mu(\zeta^a)\right] \, .
\end{equation}
Using the temporal-transverse gauge --- in which
$ X^\mu=(\tau,\xx)$ and $\xd\cdot\xp^{(i)}=0$, where $(i)=1,\cdots,p$ and ${{'(i)}}$ represents a derivative with respect to $\zeta^{(i)}$, so that $\xd$ is the velocity of the defect and the vectors $\xp^{(i)}$ represent the defect's tangents ---
one then finds that, in Fourier space~\cite{Albrecht:1997mz,Pogosian:1999np,Sousa:2015cqa}:\footnote{We found a factor of $a$ missing in eq.~(26) of ref.~\cite{Sousa:2015cqa} and that the assumption of scaling in eq.~(19) of ref.~\cite{Lazanu:2015fua} leads to a similar misrepresentation of the evolution of $\rho_{\rm DW}$. Including this correction, as we do here, results in a suppression of the CMB anisotropies compared to the predictions of~\cite{Lazanu:2015fua,Sousa:2015cqa} that is larger with increasing $\ell$. As we will show later, this leads to weaker constraints than those reported therein.}
\bq
\Theta_{00}(\mathbf{k},\tau) & = & 2^p a^{p-1}\sqrt{2}\gamma \sigma_p \cos{\left(\mathbf{k}\cdot\xx_0 + vk \tau \xhd{3} \right)}\prod_{(i)=1}^p \frac{\sin{\left(\frac{1}{2}k L_c \xhp{3}{i}\right)}}{k\xhp{3}{i}}\,,\label{eq:theta00}\\
\Theta_{ij}(\mathbf{k},\tau) & = & \Theta_{00}(\mathbf{k},\tau) \left[v^2 \xhd{i}\xhd{j}-(1-v^2)\sum_{(i)=1}^p\xhp{i}{i}\xhp{j}{i}\right]\,,\label{eq:thetaij}
\eq
where we have used the fact that
\be
\xx=\xx_0+v\tau \hat{\xd}+\sum_{(i)=1}^p \zeta^{(i)}\hat{\xp}^{(i)}\,,
\ee
with $\xx_0$ representing the position of the center of mass of the segment or section and $\hat{\xd}$ and $\hat{\xp}^{(i)}$ being unit vectors with the direction of ${\xd}$ and ${\xp}^{(i)}$, respectively (note that ${\xd}=v\hat{\xd}$ and ${\xp}^{(i)}=\zeta^{(i)}\hat{\xp}^{(i)}$). Moreover, the subscripts indicate that we are considering the projection of the vector in question along the corresponding spatial direction (e.g., $\xhd{i}=\xd\cdot \hat{\mathbf{e}}_i$). We have also assumed, without loss of generality, that $\mathbf{k}=k\hat{\mathbf{e}}_3$ and included a factor of $\sqrt{2}$ in eq.~\eqref{eq:theta00} to compensate for the fact that we are only considering the real part of $\Theta_{\mu\nu}$~\cite{Pogosian:2006hg}.

The scalar, vector, and tensor components of the anisotropic stress may then be found using
\bq
2\Theta_S & = & 2\Theta_{33}-\Theta_{11}-\Theta_{22}\,,\\
\Theta_V & = & \Theta_{13}\,,\\
\Theta_T & = & \Theta_{12}\,,
\eq
while the velocity field is given by $\Theta_D=\Theta_{03}$ for this choice of $\mathbf{k}$ and the isotropic stress $\Theta=\Theta_{ii}$ may be found by imposing local energy-momentum conservation. 

Although derived directly from the Nambu-Goto action, eqs.~\eqref{eq:theta00} and~\eqref{eq:thetaij}, with $p=1$, may also be used to describe AH strings, as their energy density is localized within a thin core. It is the choice of the VOS model that will allow us to distinguish between the two cosmic string cases, as they predict a different evolution for the string characteristic length and RMS velocity. For domain walls, we will also use the stress energy tensor in eqs.~\eqref{eq:theta00} and~\eqref{eq:thetaij} but with $p=2$, given their different dimensionality, and use the VOS model for domain walls to determine the evolution of the size of the sections and their velocity. Note that, by using the VOS model to set the length, number, and velocity of segments/sections, we ensure an accurate description of the stress-energy tensor throughout cosmic history, including the radiation-matter and matter-$\Lambda$ transitions. Notice also that the original \texttt{CMBACT} includes, for NG strings, an additional parameter $\alpha$ to account for the impact of small-scale structure on the strings. This parameter results from the effective model in~\cite{Vilenkin:1990mz,Carter:1990nb}, in which wiggly strings are modelled, on sufficiently large scales, as smooth strings with an effective tension $T_{\rm eff}$ that differs from the effective mass per unit length $\mu_{\rm eff}$:
\be 
\label{eq:alpja}
\mu_{\rm eff}=\alpha\mu\quad\mbox{and}\quad T_{\rm eff}=\mu/\alpha\,,
\ee 
whereas for strings without small-scale structure tension and mass per unit length are equivalent. The CMB anisotropies scale dominantly as $\sim (G\mu_{\rm eff})^2=(G\mu\alpha)^2$ (see e.g.~\cite{Pogosian:2006hg}).\footnote{Small-scale structure, however, should also suppress the $\Theta_{ij}$ components relative to $\Theta_{00}$~\cite{Pogosian:2006hg,Pogosian:2007gi,Avgoustidis:2012gb,Silva:2023diq}, but this effect should be subdominant. Note however that this effective description through the parameter $\alpha$, by treating strings as being effectively smooth, may be misrepresenting this effect~\cite{Silva:2023diq}.} Throughout this work we will then set $\alpha=1$, as this provides the most conservative constraints on $G\mu$.\footnote{The constraints for other values of $\alpha$ may, however, be roughly estimated by rescaling by $\alpha$.}

\section{CMB anisotropies in the presence of defect networks}
\label{sec:CMB-anisotropies}
The scalar, vector and tensor components of the energy-momentum tensor of topological defects act as additional source terms, $\Theta_{\mu\nu}$, in the linearized Einstein equations for the metric perturbations, $\delta G_{\mu\nu} = 8\pi G \left( \delta T_{\mu\nu} + \Theta_{\mu\nu} \right)$, where $\delta T_{\mu\nu}$ corresponds to the contributions from the standard cosmological components. Since metric perturbations directly enter the Boltzmann equations for photons and matter perturbations, topological defects consequently have an impact on the evolution of primordial fluctuations. Unlike inflationary scenarios, where perturbations are set by initial conditions and evolve passively after horizon re-entry, defects serve as active and incoherent gravitational sources that continuously generate perturbations throughout cosmic history. This distinctive mechanism leads to characteristic signatures in the CMB anisotropies, notably the suppression of acoustic peaks in the power spectra generated by the networks~\cite{Albrecht:1995bg,Albrecht:1997mz,Durrer:2001cg}.

The modified Einstein equations in the presence of cosmic strings are implemented in the \texttt{CMBACT4} code~\cite{Pogosian:1999np,Pogosian:2006hg}, which computes the CMB spectra generated by string networks based on the phenomenological USM described in the previous sections. 
For domain walls, we use a modified version of the same code, previously employed in~\cite{Sousa:2015cqa}.
The defect-induced spectra are obtained by averaging over 2500 independent realizations of the network and are computed assuming $L_f = 0.5$ in eq.~\eqref{eq:Toff}. 
Adopting $L_f = 0.5$ leads to a reduced defect contribution to the CMB power spectra and therefore to more conservative constraints on the string and domain wall tension.
The dependence of the results on the choice of $L_f$, including the corresponding constraints obtained for $L_f = 1$, is discussed in section~\ref{sec:results} (see also appendix~\ref{appendix:triangle_plots} for more details). The resulting defect-induced spectra are then added to the inflationary ones computed with \texttt{CAMB}~\cite{Lewis:1999bs} according to
\begin{equation}
    \label{eq:total-spectra}
    C_\ell = C_\ell^{\Lambda{\rm CDM}} + A_{\rm defect} C_\ell^{\rm defect} \, ,
\end{equation}
where $C_\ell^{\rm defect}$ denotes the defect spectra computed for a reference tension. 
The dimensionless parameter $A_{\rm defect}$ rescales the defect spectra according to the actual network tension and controls the relative contribution of defects to the total CMB anisotropies. 
This is defined as
\begin{equation}
    \label{eq:A-defect}
    A_{\rm str} \equiv \left(\frac{G\mu}{10^{-6}}\right)^2 \, , 
    \qquad 
    A_{\rm DW} \equiv \left(\frac{\sigma^{1/3}}{1 \; {\rm MeV}}\right)^6 \, ,
\end{equation}
for strings and domain walls, respectively.

\begin{figure}[!t]
    \centering
    \includegraphics[width=1\linewidth]{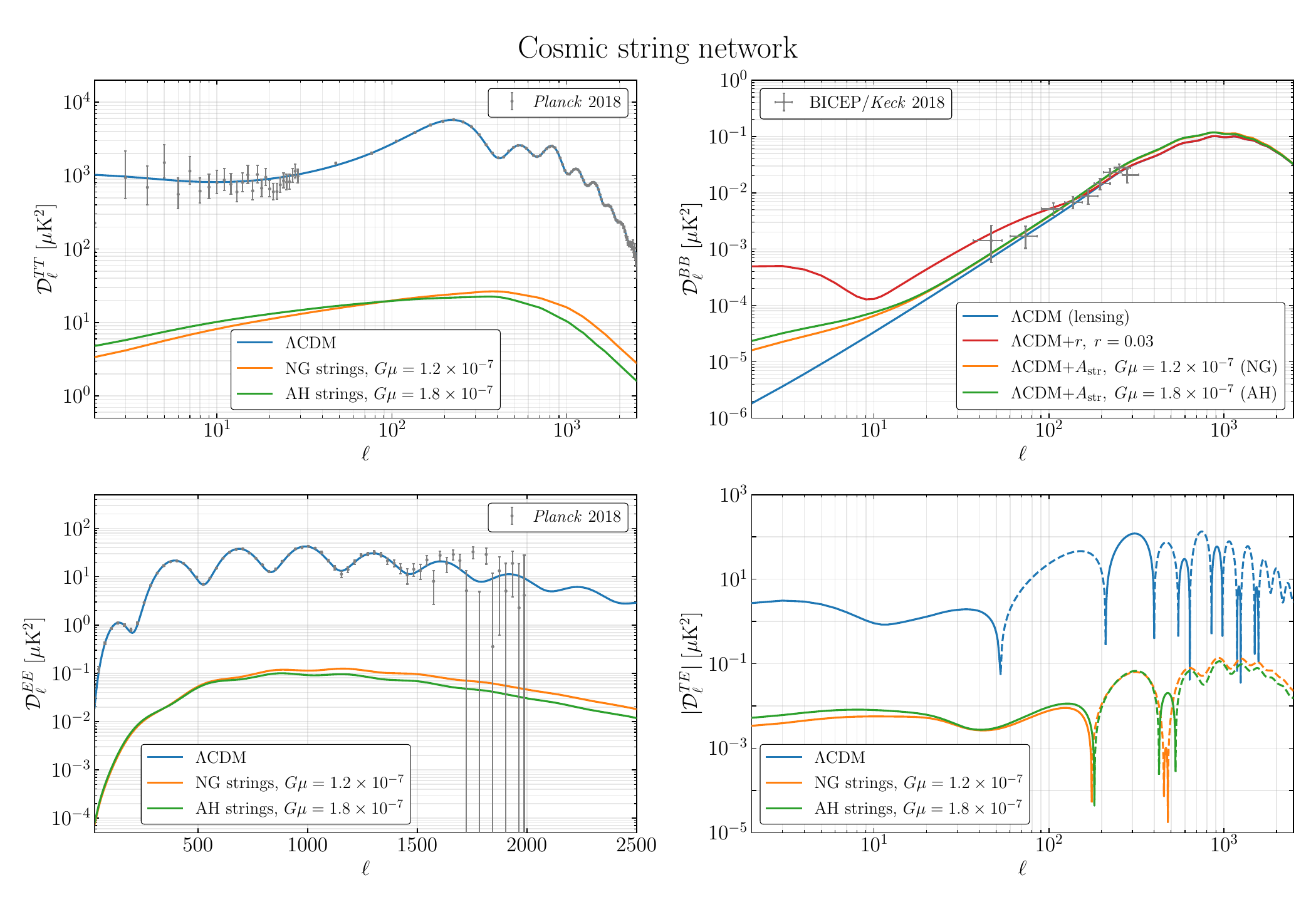}
    \caption{CMB angular power spectra sourced by cosmic string networks compared to the $\Lambda$CDM prediction and observational data. 
    {\it Top left}: Temperature power spectrum, $\mathcal{D}_\ell^{TT} \equiv \ell (\ell + 1) C_\ell^{TT} / (2\pi)$, for NG (orange) and AH (green) strings, together with the $\Lambda$CDM prediction (blue). The $\Lambda$CDM parameters are fixed to the values inferred from \textit{Planck} 2018 data. 
    {\it Top right}: $B$-mode power spectrum, $\mathcal{D}_\ell^{BB}$, for the baseline $\Lambda$CDM model (blue), $\Lambda$CDM+$r$ with $r=0.03$ (red), and $\Lambda$CDM plus NG and AH strings.
    {\it Bottom left}: $E$-mode power spectrum, $\mathcal{D}_\ell^{EE}$, shown for $\ell \ge 30$. 
    {\it Bottom right}: Absolute value of the $TE$ power spectrum, $|\mathcal{D}_\ell^{TE}|$. Solid segments correspond to $\mathcal{D}_\ell^{TE}>0$, while dashed segments indicate $\mathcal{D}_\ell^{TE}<0$.
    In all panels, the string spectra are computed for tensions corresponding to the 95\% credible upper limits derived from the \textit{Planck}$+$BK18 analysis, namely $G\mu = 1.2 \times 10^{-7}$ for NG strings and $G\mu = 1.8 \times 10^{-7}$ for AH strings. 
    Measurements from \textit{Planck} 2018 (gray points with error bars)~\cite{Planck:2019nip} are shown for the $TT$ and $EE$ spectra, while $B$-mode data from BICEP/\textit{Keck} 2018~\cite{BICEP:2021xfz} are included in the top right panel.}
    \label{fig:Spectra_strings}
\end{figure}

Figure~\ref{fig:Spectra_strings} illustrates the impact of cosmic string networks on the CMB power spectra, for both NG and AH strings. The upper-left panel shows the temperature power spectra, $\mathcal{D}_\ell^{TT} \equiv \ell (\ell + 1) C_\ell^{TT} / (2\pi)$, sourced by cosmic string networks alone, i.e. without adding the standard $\Lambda$CDM contribution.
These are shown for string tensions of $G\mu = 1.2 \times 10^{-7}$ for NG strings and $G\mu = 1.8 \times 10^{-7}$ for AH strings, corresponding to the 95\% credible upper limits derived from our \textit{Planck}$+$BK18 analysis (see section~\ref{sec:results}).
The $\Lambda$CDM spectrum is shown for reference, to highlight the relative amplitude and shape of the string-induced signal.
The temperature spectrum sourced by cosmic strings is smooth, lacks acoustic oscillations, and exhibits a broad peak at $\ell \sim 350$.
In the case of NG strings the peak is slightly shifted towards smaller angular scales with respect to the AH case, reflecting the smaller characteristic length-scale of the NG network (see the left panel of figure~\ref{fig:VOSEvolution}), which provides a measure of the string correlation length. This difference originates from the distinct energy-loss mechanisms operating in the two string models.
For the same reason, at fixed string tension, the overall amplitude of the signal sourced by AH strings is suppressed with respect to the NG case.

In contrast, the upper-right panel shows the total $B$-mode power spectra, $\mathcal{D}_\ell^{BB}$, including the $\Lambda$CDM contribution from lensing of $E$-modes.
For comparison, we also show the $\Lambda$CDM$+r$ prediction from inflationary tensor modes, with $r=0.03$, together with the BICEP/\textit{Keck} 2018 (BK18) measurements~\cite{BICEP:2021xfz}.
The comparison shown here assumes a tensor-to-scalar ratio $r = 0.03$, which is within the 95\% credible upper limit obtained from BK18~\cite{BICEP:2021xfz} (see~\cite{Galloni:2022mok} for a stronger bound on $r$ obtained including additional datasets). Note in particular that at multipoles $\ell \gtrsim 100$ the $B$-mode signal sourced by cosmic strings exceeds the contribution from inflationary tensor modes.

The lower panels show the $EE$ (left) and $TE$ (right) spectra sourced by cosmic strings. As in the $TT$ case, acoustic oscillations are suppressed, reflecting the incoherent and active nature of the perturbations sourced by cosmic string networks.

\begin{figure}
    \centering
    \includegraphics[width=1\linewidth]{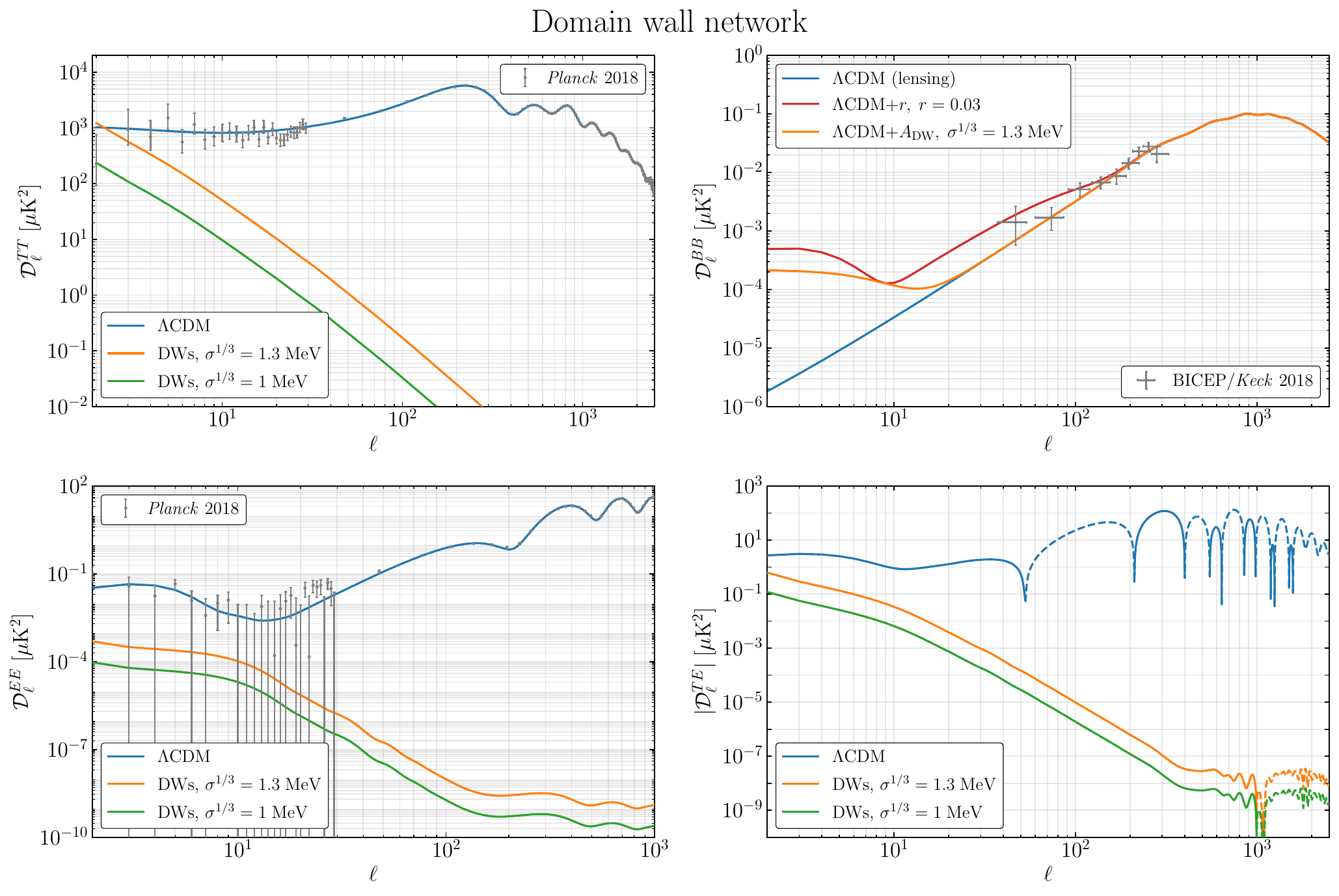}
    \caption{CMB angular power spectra for models including a domain wall network compared to the $\Lambda$CDM prediction and observational data.
    {\it Top left}: Temperature power spectrum, $\mathcal{D}_\ell^{TT} \equiv \ell (\ell + 1) C_\ell^{TT} / (2\pi)$, for the baseline $\Lambda$CDM model (blue), together with the contributions sourced by DW networks alone for surface tensions $\sigma^{1/3} = 1.3 \; \mathrm{MeV}$ (orange), corresponding to the 95\% credible upper limit derived in this work (see section~\ref{sec:results}), and $\sigma^{1/3} = 1 \; \mathrm{MeV}$ (green), corresponding to the original Zel'dovich-Kobzarev-Okun bound~\cite{Zeldovich:1974uw}. The same colour coding is adopted in all panels.
    {\it Top right}: $B$-mode power spectrum, $\mathcal{D}_\ell^{BB}$, for the baseline $\Lambda$CDM model (blue), $\Lambda$CDM+$r$ with $r=0.03$ (red), and $\Lambda$CDM plus a DW network with $\sigma^{1/3} = 1.3 \; \mathrm{MeV}$ (orange).
    {\it Bottom left}: $E$-mode power spectrum, $\mathcal{D}_\ell^{EE}$, shown for $2 \le \ell \le 1000$.
    {\it Bottom right}: $TE$ cross-correlation power spectrum, $\mathcal{D}_\ell^{TE}$.
    Measurements from \textit{Planck} 2018 (gray points with error bars)~\cite{Planck:2019nip} are shown for the $TT$, $EE$, and $TE$ spectra, while $B$-mode data from BICEP/\textit{Keck} 2018~\cite{BICEP:2021xfz} are included in the top right panel.}
    \label{fig:Dl_TT_stable_data}
\end{figure}

We now turn to the CMB signatures of DW networks, which are illustrated in figure~\ref{fig:Dl_TT_stable_data}.
We consider two different values of the DW tension, $\sigma^{1/3} = 1 \; {\rm MeV}$ and $\sigma^{1/3} = 1.3 \; {\rm MeV}$.
The first value corresponds to the original Zel'dovich-Kobzarev-Okun bound for stable networks~\cite{Zeldovich:1974uw}, later refined in~\cite{Sousa:2015cqa,Lazanu:2015fua,Ferreira:2023jbu}.
The second value represents the 95\% credible upper limit derived in this work from a full MCMC analysis (see section~\ref{sec:results}).

In the top left panel, we show the temperature power spectra, $\mathcal{D}_\ell^{TT}$, sourced by DW networks alone together with the $\Lambda$CDM prediction, compared to \textit{Planck} 2018 data~\cite{Planck:2019nip}.
As expected, stable DW networks contribute at the largest scales, as their energy density dilutes more slowly than that of radiation and matter, leading to an enhanced contribution at late times.

In the top right panel, we show the total $B$-mode power spectrum for $\sigma^{1/3} = 1.3 \; {\rm MeV}$.
Compared to the $\Lambda$CDM case with no primordial tensor modes (only lensing), the inclusion of DWs leads to increased power at large angular scales, due to the contribution from vector and tensor perturbations sourced by the DW network (see also~\cite{Sousa:2015cqa}).
Even when inflationary tensor modes are included (i.e., $r \ne 0$), the two signals remain in principle distinguishable at large and intermediate scales ($\ell \lesssim 100$).

The bottom panels show the $E$-mode and $TE$ cross-correlation power spectra, where the DW contribution is again concentrated at large angular scales.


\section{Constraints from \textit{Planck} 2018 and BK18 data}
\label{sec:constraints}
In this section, we present the constraints on cosmic string and domain wall networks obtained using data from {\it Planck} 2018 and BICEP/{\it Keck} 2018. We first describe the datasets and the analysis methodology, and then discuss the resulting bounds and their physical implications.

\subsection{MCMC analysis: methodology and datasets}
\label{sec:analysis}
We perform a Markov Chain Monte Carlo (MCMC) analysis to constrain the $\Lambda$CDM (or $\Lambda$CDM$+r$) model extended by the presence of a network of cosmic strings or domain walls. 
As discussed in section~\ref{sec:CMB-anisotropies}, the CMB spectra sourced by defects are computed using \texttt{CMBACT4}~\cite{Pogosian:1999np,Pogosian:2006hg}, keeping the cosmological parameters fixed to the values obtained in the \textit{Planck} 2018 analysis in combination with baryon acoustic oscillation data~\cite{Planck:2018vyg}. These spectra, rescaled by the dimensionless wall and string amplitudes, respectively $A_{\rm DW}$ or $A_{\rm str}$, are added to the inflationary spectra obtained with \texttt{CAMB}~\cite{Lewis:1999bs} (see eqs.~\eqref{eq:total-spectra} and~\eqref{eq:A-defect}).\footnote{This approach is commonly adopted when deriving CMB constraints on defects (see, e.g., refs.~\cite{Planck:2015fie,Charnock:2016nzm} for cosmic strings). Indeed, computing the defect spectra at each step of the MCMC, varying also the $\Lambda$CDM parameters, would be computationally prohibitive. Since defects contribute at the percent level to the total $C_\ell$~\cite{Planck:2015fie}, and their spectra are only mildly sensitive to variations of the $\Lambda$CDM parameters within a few standard deviations from the values inferred from CMB data, keeping these parameters fixed in the computation of the defect spectra has a minor impact on the resulting constraints.}
The MCMC analysis is then performed using \texttt{Cobaya}~\cite{Torrado:2020dgo} over the parameter set 
\begin{equation}
    \{ \omega_b,\, \omega_c,\, 100 \, \theta_{\rm MC}, \, \tau_{\rm reio}, \, \ln(10^{10}A_s), \, n_s,\, r, \, A_{\rm defect} \} \, ,    
\end{equation}
where $\omega_b \equiv \Omega_b h^2$ and $\omega_c \equiv \Omega_c h^2$ denote the physical density of baryons and cold dark matter, respectively, $\theta_{\rm MC}$ parametrizes the angular size of the acoustic scale at recombination, $\tau_{\rm reio}$ is the reionization optical depth, $A_s$ is the amplitude of the primordial power spectrum at the pivot scale $k_* = 0.05 \; {\rm Mpc}^{-1}$, $n_s$ is the scalar spectral index and $r$ is the tensor-to-scalar ratio. We consider a spatially flat Universe with adiabatic initial conditions. Neutrinos are modeled with one massive and two massless species, and the sum of neutrino masses is fixed to the minimal value allowed by flavor oscillation experiments in the normal hierarchy scenario, $\sum m_\nu = 0.06 \; {\rm eV}$~\cite{deSalas:2020pgw,Esteban:2020cvm,Capozzi:2021fjo}. The effective number of relativistic species is set to its standard value, $N_{\rm eff} = 3.044$~\cite{Mangano:2001iu,Bennett:2019ewm,Bennett:2020zkv,Akita:2020szl,Froustey:2020mcq,Cielo:2023bqp,Drewes:2024wbw}.
Details on the priors adopted for the parameters varied in our analysis are provided in table~\ref{tab:priors}.
\begin{table}[!t]
    \centering
    \resizebox{\textwidth}{!}{
    \begin{tabular}{ccccccccccc}
        \hline      
        \hline 
        \noalign{\vskip 1mm}
        Parameter & $\omega_b$ & $\omega_c$ & $100\,\theta_{\rm MC}$ & $\tau_{\rm reio}$ & $\ln(10^{10}A_s)$ & $n_s$ & $r$ & $A_{\rm DW}$ & $A_{\rm str}$ \\
        \noalign{\vskip 1mm}
        \hline
        \noalign{\vskip 1mm}
        Prior & $[0.005, 0.1]$ & $[0.001, 0.99]$ & $[0.5, 10]$ & $[0.01, 0.8]$ & $[1.61, 3.91]$ & $[0.8, 1.2]$ & $[0, 1]$ & $[0, 15]$ & $[0, 3]$ \\[1mm]
        \hline
        \hline 
    \end{tabular}
    }
    \caption{Uniform priors adopted for the parameters varied in the MCMC analysis.}
    \label{tab:priors}
\end{table}
In particular, we use flat priors on the defect amplitudes $A_{\rm str}$ and $A_{\rm DW}$. 
An alternative possibility is to assume a flat prior on the logarithm of the defect tension, as done, e.g., in~\cite{Charnock:2016nzm,Battye:2006pk,Battye:2010xz} for cosmic strings. In the case of domain walls, we have explicitly verified that this choice leads to tighter upper limits with respect to a flat prior on $A_{\rm DW}$, in agreement with what was found in~\cite{Battye:2010xz} for cosmic strings. This difference originates from prior-volume effects in the regime where the defect contribution to the CMB power spectra is small: a flat prior on the logarithm of the tension assigns comparatively more weight to values yielding CMB spectra indistinguishable from $\Lambda$CDM, thereby strengthening the resulting constraints.\footnote{In practice, this also results in a less efficient exploration of the posterior distribution, since a large fraction of the sampling time is spent probing regions where the posterior is nearly flat over several orders of magnitude in the tension, while the region where the posterior rapidly decreases, crucial for determining the upper bound, is explored less efficiently.}

To ensure convergence of the chains, we require the Gelman-Rubin parameter to satisfy $R - 1 \lesssim 0.01$~\cite{Gelman:1992zz}, after discarding the burn-in. The posterior distributions are then obtained and analyzed using the \texttt{GetDist} package~\cite{Lewis:2019xzd}.

The datasets used in our analysis are:
\begin{itemize}
    \item \textit{Planck}: large-scale ($2 \le \ell \le 29$) temperature and $E$-mode polarization data from the \texttt{Commander}~\cite{Planck:2018vyg,Planck:2019nip} and \texttt{SRoll2}~\cite{Pagano:2019tci,Delouis:2019bub} likelihoods, respectively, complemented at high multipoles ($\ell \ge 30)$ by the \texttt{Plik\_lite} likelihood~\cite{Planck:2018vyg,Planck:2019nip}.\footnote{For cosmic strings, we have also performed a run using the full multi-frequency \texttt{Plik} likelihood at $\ell \ge 30$~\cite{Planck:2019nip}. The resulting constraints are fully consistent with those obtained using the \texttt{Plik\_lite} likelihood.} We also include CMB lensing measurements from {\it Planck} PR4~\cite{Carron:2022eyg}.\footnote{The lensing contribution sourced by defect-induced perturbations, as well as the lensing applied to the defect CMB spectra, is neglected in our analysis. Given that topological defects contribute at most at the percent level to the total CMB and matter power spectra, the corresponding lensing corrections are expected to be subdominant. \textit{Planck} lensing data are thus used only to improve constraints on $\Lambda$CDM parameters.}
    \item BICEP/\textit{Keck} 2018 (BK18): $B$-modes data collected by the BICEP2, \textit{Keck Array} and BICEP3 experiments up to the 2018 observing season~\cite{BICEP:2021xfz}.
\end{itemize}


\subsection{Results and discussion}
\label{sec:results}
The main results of our analysis are summarized in table~\ref{tab:constraints_datasets}, which reports the 95\% credible limits on the string and domain wall tensions, together with the bound on the tensor-to-scalar ratio $r$, for the different models and data combinations considered in this work.
For completeness, the full set of constraints obtained from the \textit{Planck}$+$BK18 analysis is reported in table~\ref{tab:constraints}.

\begin{table}[!t]
    \centering
    \resizebox{1\textwidth}{!}{
    \begin{tabular}{ c c c c c }
        \hline
        \hline
        \noalign{\vskip 1mm}
        Model 
        & 
        & \textit{Planck} ($r$ fixed to zero)
        & \textit{Planck}$+$BK18 ($r$ fixed to zero)
        & \textit{Planck}$+$BK18 ($r$ varied) \\
        \noalign{\vskip 1mm}
        \hline
        \noalign{\vskip 1mm}
        \multirow{3}{*}{NG strings} 
        & $A_{\rm str}$ 
        & $< 0.018$ & $< 0.015$ & $< 0.014$ \\[1mm]
        & $G\mu$ 
        & $< 1.3 \times 10^{-7}$ & $< 1.2 \times 10^{-7}$ & $< 1.2 \times 10^{-7}$ \\[1mm]
        & $r$
        & -- & -- & $< 0.031$ \\[1mm]
        \hline
        \noalign{\vskip 1mm}
        \multirow{3}{*}{AH strings} 
        & $A_{\rm str}$ 
        & $< 0.049$ & $< 0.033$ & $< 0.031$ \\[1mm]
        & $G\mu$ 
        & $< 2.2 \times 10^{-7}$ & $< 1.8 \times 10^{-7}$ & $< 1.8 \times 10^{-7}$ \\[1mm]
        & $r$
        & -- & -- & $ < 0.030$ \\[1mm]
        \hline
        \noalign{\vskip 1mm}
        \multirow{3}{*}{Domain walls} 
        & $A_{\rm DW}$ 
        & $< 5.4$ & $< 5.2 $ & $< 5.2$ \\[1mm]
        & $\sigma^{1/3}\;[\mathrm{MeV}]$ 
        & $< 1.3$ & $< 1.3$ & $< 1.3$ \\[1mm]
        & $r$
        & -- & -- & $< 0.035 $ \\[1mm]
        \hline
        \hline
    \end{tabular}
    }
\caption{95\% credible upper limits on $A_{\rm str}$ and $A_{\rm DW}$, together with the corresponding derived limits on the string tension $G\mu$ and the domain wall tension $\sigma^{1/3}$. In the last column the tensor-to-scalar ratio $r$ is varied together with the string/domain wall tension and the six $\Lambda$CDM parameters. For reference, the baseline \textit{Planck}$+$BK18 analysis without defects yields an upper bound $r < 0.035$ (see table~\ref{tab:constraints}).}
\label{tab:constraints_datasets}
\end{table}

\begin{figure}[!t]
    \centering
    \includegraphics[width=0.49\linewidth]{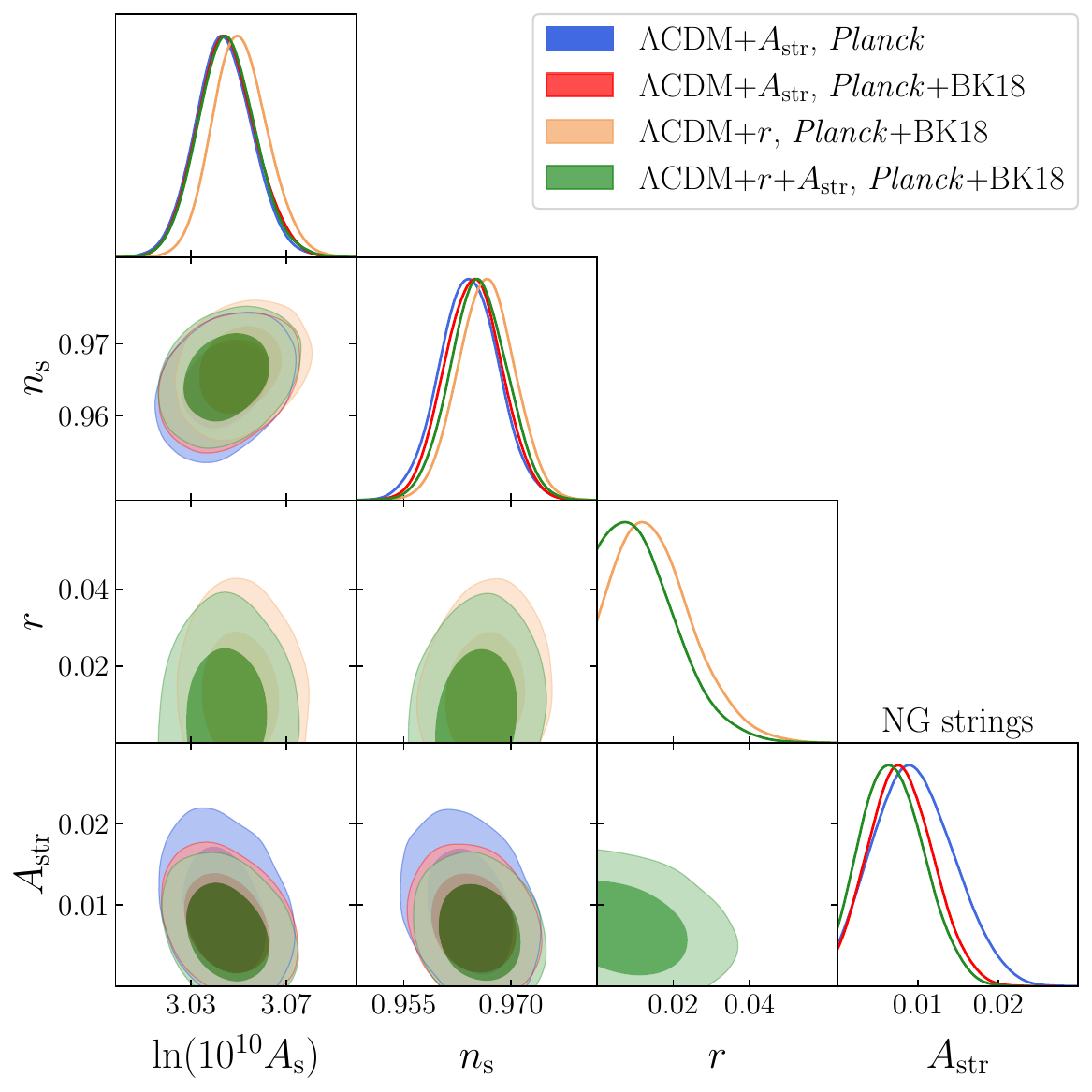}
    \includegraphics[width=0.49\linewidth]{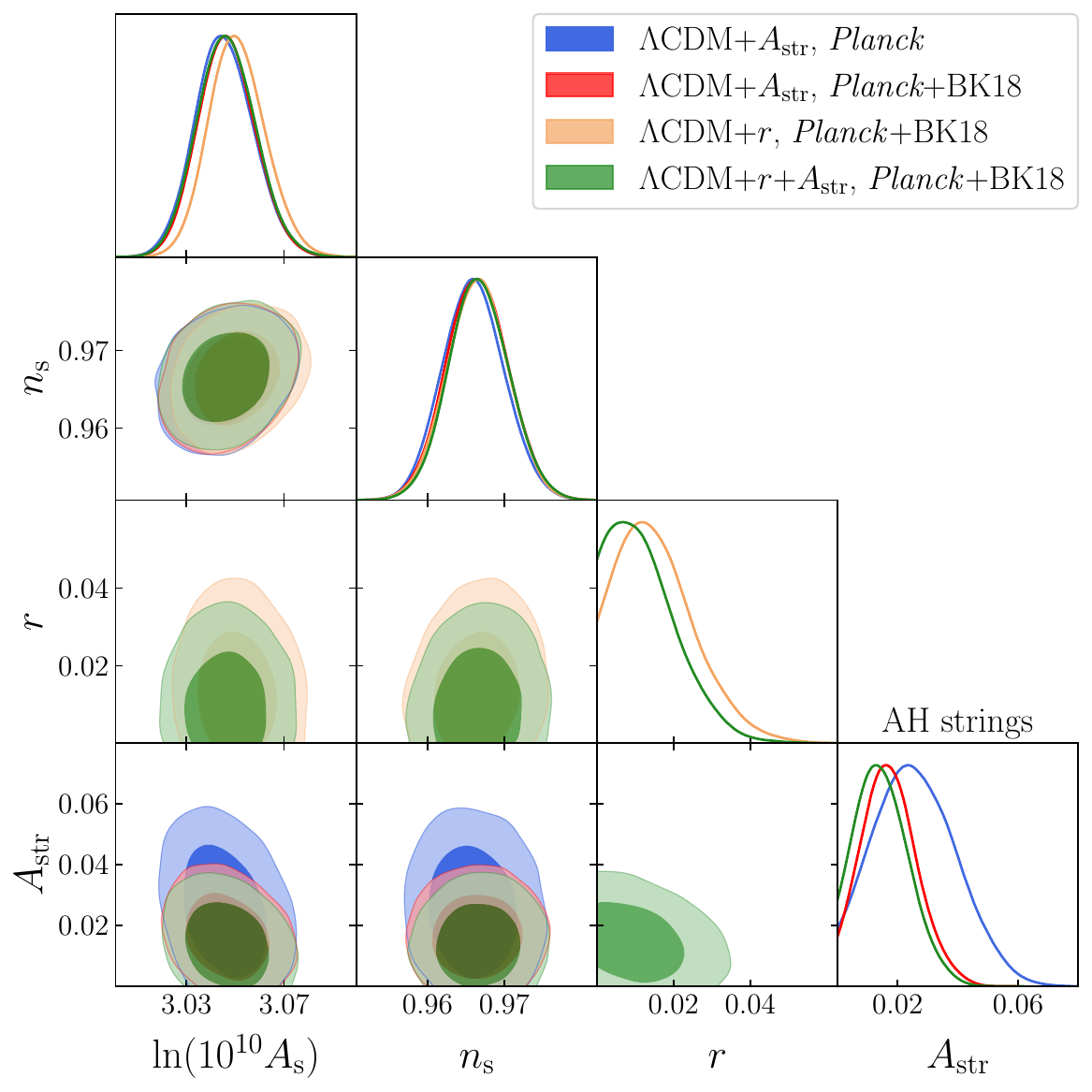}
    \caption{One and two-dimensional posterior distribution functions for $A_{\rm str}$ and the cosmological parameters that exhibit the strongest correlations with the string tension, for NG strings (left) and AH strings (right). The complete triangle plots including all the $\Lambda$CDM parameters are shown in appendix~\ref{appendix:triangle_plots} (see figures~\ref{fig:Strings_NG_triangle_full} and~\ref{fig:Strings_AH_triangle_full}).}
    \label{fig:Strings_NG_triangle}
\end{figure}

We start by discussing the results obtained for cosmic string networks.
Using the \textit{Planck}$+$BK18 dataset, we obtain 95\% credible upper limits on the string tension of $G\mu < 1.2 \times 10^{-7}$ for NG strings and $G\mu < 1.8 \times 10^{-7}$ for AH strings. Compared to constraints derived from \textit{Planck} data alone, the inclusion of BK18 leads to an
improvement of about $8\%$ for NG strings and $18\%$ for AH strings.
This larger improvement for AH strings can be understood as follows. As discussed in section~\ref{sec:CMB-anisotropies}, at fixed string tension the overall amplitude of the signal sourced by AH strings is suppressed with respect to the NG case. However, this suppression is more pronounced in temperature than in $B$-mode polarization, making the $B$-mode peak relatively more pronounced for AH strings. As a consequence, $B$-mode measurements have a stronger impact on the constraints for this model.
As both cosmic strings and primordial tensor modes add power to the $B$-mode spectrum, we also find a mild anti-correlation between $A_{\rm str}$ and $r$, resulting in stronger bounds on the two parameters when they are both allowed to vary (see figure~\ref{fig:Strings_NG_triangle}). We note, however, that once the constraints are translated into limits on the string tension $G\mu$ and rounded to the quoted significant digits, the bounds obtained with fixed and varying $r$ are effectively identical, as a consequence of the quadratic scaling $A_{\rm str}\propto (G\mu)^2$. 

From the posterior distributions shown in figure~\ref{fig:Strings_NG_triangle}, we also observe mild anti-correlations between $A_{\rm str}$ and the $\Lambda$CDM parameters $A_s$ and $n_s$ (see figures~\ref{fig:Strings_NG_triangle_full} and~\ref{fig:Strings_AH_triangle_full} for the full triangle plots).
These reflect the fact that string networks enhance the CMB power at intermediate to small angular scales, an effect that can be compensated either by decreasing the amplitude of the primordial power spectrum $A_s$, or by shifting the spectrum further away from scale invariance, i.e. lowering $n_s$.

Our constraints for both NG and AH strings are stronger than previous results based on {\it Planck} 2015 data~\cite{Planck:2015fie,Charnock:2016nzm,Lizarraga:2016onn}. Although an exact one-to-one comparison is not possible due to the differences in the implementation of the model,\footnote{For example, in~\cite{Charnock:2016nzm} a stronger constraint on $G\mu$ is found for NG strings. However, as the authors note, that is due to the use of $L_f=1$ rather than our conservative choice of $L_f=0.5$. As we show in figure~\ref{fig:Strings_NG_triangle_full} (see also table~\ref{tab:constraints_Lf1}), we also find stronger constraints when using $L_f=1$. For AH strings, instead, the constraints obtained in~\cite{Planck:2015fie,Lizarraga:2016onn} do not rely on the USM but instead on the UETC taken directly from field theory simulations.} 
the inclusion of new data, and in particular $B$-mode measurements from BK18, leads to an overall improvement of about $20-30\%$ at the level of the power spectra, or equivalently, $10-20\%$ at the level of the string tension.\footnote{We have also checked that the inclusion of high-$\ell$ $EE$ and $TE$ data improves the constraint on the string tension by approximately 20\% with respect to temperature and low-$\ell$ polarization alone.}

\begin{table}[!t]
    \centering
    \resizebox{0.6\textwidth}{!}{
    \begin{tabular}{ c c c } 
        & \multicolumn{2}{c}{$\Delta\chi^2$ (preference)} \\
        \hline
        \hline
        \noalign{\vskip 1mm}
        Data/Model & NG strings & AH strings \\
        \noalign{\vskip 1mm}
        \hline
        \noalign{\vskip 1mm}
        {\it Planck} with $r$ fixed
        & 2.81 ($1.68\sigma$) & 3.15 ($1.77\sigma$) \\[1mm]
        {\it Planck}$+$BK18 with $r$ fixed 
        & 1.80 ($1.34\sigma$) & 1.04 ($1.02\sigma$) \\[1mm]
        {\it Planck}$+$BK18 with $r$ varied 
        & 2.19 ($1.48\sigma$) & 1.18 ($1.09\sigma$) \\[1mm]
        \hline
        \hline
    \end{tabular}
    }
    \caption{Goodness-of-fit, $\Delta\chi^2 \equiv \chi^2_{\rm baseline}-\chi^2_{\rm str}$, for Nambu-Goto and Abelian-Higgs strings, relative to the baseline $\Lambda$CDM model, or to $\Lambda$CDM$+r$ when $r$ is varied together with $A_{\rm str}$. In parentheses, we report the corresponding preference for string models over the baseline one.}
    \label{tab:deltachi2_strings}
\end{table}

Interestingly, the posterior distributions in figure~\ref{fig:Strings_NG_triangle} show a mild preference for a non-zero cosmic string tension in both the NG and AH string cases, with the corresponding best-fit values reported in parentheses in table~\ref{tab:constraints}. To quantify the preference for models with strings over the baseline $\Lambda$CDM (or $\Lambda$CDM$+r$, when $r$ is varied), we consider the improvement in the best-fit $\chi^2$, defined as $\Delta\chi^2 \equiv \chi^2_{\rm baseline}-\chi^2_{\rm str}$. Assuming that the $\Delta\chi^2$ follows a $\chi^2$ distribution with one degree of freedom, the corresponding preference is then estimated as $\sqrt{\Delta\chi^2}$ in units of $\sigma$.
The values of $\Delta\chi^2$ and the associated preference levels for the different datasets and models are reported in table~\ref{tab:deltachi2_strings}, with the latter lying between $1.02\sigma$ and $1.77 \sigma$.
With \textit{Planck} data alone, the preference is similar for NG and AH strings, and slightly larger for the latter. The inclusion of BK18 data lowers the preference in both models, with a larger reduction for AH strings. This follows from the stronger constraining power of $B$-mode measurements for this string model.

As discussed in section~\ref{sec:CMB-anisotropies}, the constraints reported above are derived assuming $L_f = 0.5$ in the USM.
We have verified that adopting $L_f=1$ leads to systematically stronger constraints. In particular, the resulting 95\% credible upper limits on the string tension tighten by about $17\%$ for NG strings and $11\%$ for AH strings (see table~\ref{tab:constraints_Lf1}).

Finally, we note that including the contributions from string loops has a very small impact on the final constraints. This can be seen in figure~\ref{fig:Strings_NG_loops_1D}, which compares the posterior distributions of $A_{\rm str}$ for NG strings with and without loops, obtained using {\it Planck} data. Since loops give contributions at small scales, where no $B$-mode measurements are currently available, we do not expect any improvement from the inclusion of BK18 data.

We now discuss the results obtained for DW networks.
Using the \textit{Planck}$+$BK18 dataset, our analysis yields a 95\% credible upper limit on the domain wall tension of $\sigma^{1/3} < 1.3\;\mathrm{MeV}$.
This bound is weaker than those reported in previous analyses~\cite{Sousa:2015cqa,Lazanu:2015fua} for the reason discussed in footnote 10 of section~\ref{subsec:USM}.

\begin{figure}[!t]
    \centering
    \includegraphics[width=0.5\linewidth]{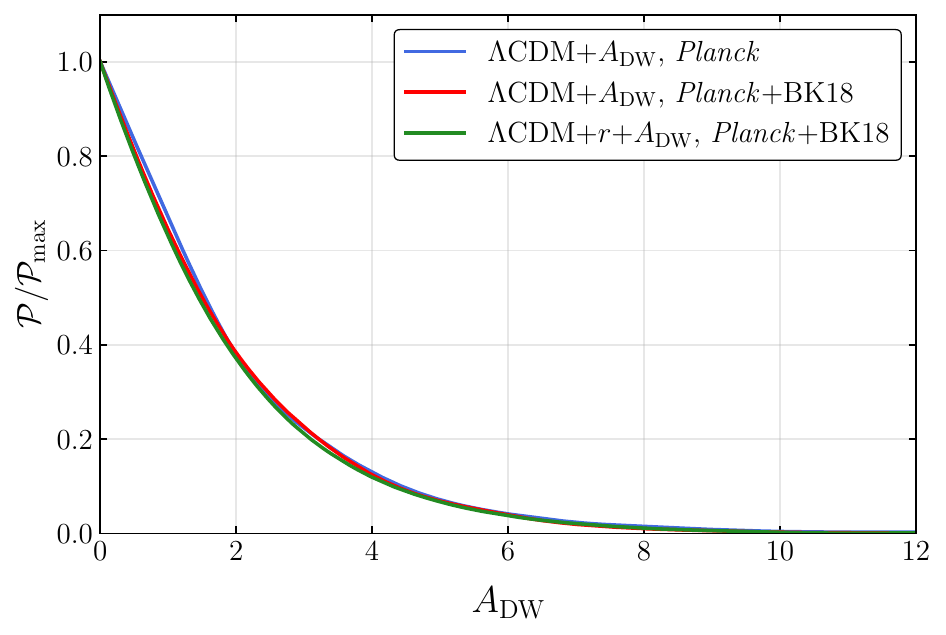}
    \caption{One-dimensional posterior distribution of $A_{\rm DW}$, for different models and dataset combinations.}
    \label{fig:DWs_1D_posteriors}
\end{figure}

In contrast to the case of cosmic strings, the inclusion of BK18 data does not lead to a significant improvement in the constraints on $A_{\rm DW}$, as can be seen from the one-dimensional posterior distributions of $A_{\rm DW}$ shown in figure~\ref{fig:DWs_1D_posteriors} (see figure~\ref{fig:DWs_stable_triangle} for the full triangle plot including all $\Lambda$CDM parameters and the tensor-to-scalar ratio $r$).
This behavior can be understood as follows.
The $B$-mode signal sourced by DWs peaks at very large scales ($\ell \lesssim 10$), well below the lowest multipole probed by BK18.
At $\ell \gtrsim 50$, where BK18 measurements start, the \textit{Planck} bound on $A_{\rm DW}$ already ensures that the DW-induced $B$-mode power remains subdominant with respect to the lensing contribution (see the right panel of figure~\ref{fig:Dl_TT_stable_data}).
As a result, BK18 does not provide additional constraining power for stable DW networks. For the same reason, no anti-correlation is observed with the tensor-to-scalar ratio $r$. We also find no appreciable correlations between $A_{\rm DW}$ and any of the $\Lambda$CDM parameters (see figure~\ref{fig:DWs_stable_triangle}), reflecting the fact that DWs affect the CMB power spectra only at very large angular scales.

As for cosmic strings, we have also verified that adopting $L_f = 1$ in the USM leads to slightly stronger constraints for DW networks. In this case, the 95\% credible upper limit on $A_{\rm DW}$ tightens by about $17\%$ with respect to the baseline choice $L_f = 0.5$ (see table~\ref{tab:constraints_Lf1}).

\section{Forecasts for future CMB surveys}
\label{sec:forecasts}

In this section we assess the expected sensitivity of upcoming CMB experiments to the presence of cosmic defect networks. 
As discussed in section~\ref{sec:CMB-anisotropies}, cosmic strings primarily affect intermediate and small angular scales, while stable domain wall networks impact only the largest angular scales. This motivates considering complementary experimental configurations sensitive to these distinct signatures.

For cosmic strings, we provide forecasts for the Simons Observatory (SO)~\cite{SimonsObservatory:2018koc} in combination with \textit{Planck}. 
For stable domain walls, we present forecasts for the \textit{LiteBIRD} satellite~\cite{LiteBIRD:2022cnt}, whose improved sensitivity to large-scale polarization, including $B$-modes, is particularly relevant for testing this scenario. Details of the experimental specifications adopted in our analysis are given in section~\ref{sec:forecasts_noise}, while the resulting projected constraints are presented and discussed in section~\ref{sec:forecasts_results}.

\subsection{Noise modeling and simulated data}
\label{sec:forecasts_noise}
To model the instrumental noise for the experiments considered in this analysis, we proceed as follows. 
When noise curves are not publicly available, we compute them from the reported instrumental specifications, assuming white noise with a Gaussian beam (see e.g.~\cite{Perotto:2006rj,Wu:2014hta,Gerbino:2019okg}):
\begin{equation}
    \label{eq:noise}
    N_\ell^{XX} = \sigma_X^2 \exp\left[ \frac{\ell(\ell+1) \theta_{\rm FWHM}^2}{8 \ln 2} \right] \, , 
\end{equation}
with $X \in \{T, E, B \}$. 
Here, $\sigma_X$ denotes the instrumental sensitivity in $\mu$K-arcmin and $\theta_{\rm FWHM}$ is the full-width half-maximum beam size, expressed in radians. 
We consider two experimental configurations:
\begin{itemize}
    \item {\it Planck}$+$SO: We model the noise properties of SO following the specifications provided by the collaboration~\cite{SimonsObservatory:2018koc}, and combine them with an effective description of \textit{Planck} noise to build the experimental configuration used in our forecasts.
    For SO, we include both the Small Aperture Telescopes (SATs), which target $B$-modes at intermediate angular scales, and the Large Aperture Telescope (LAT), which will provide high-precision measurements of temperature and $E$-mode polarization at small scales.
    We combine these experiments as follows (see e.g.~\cite{Raffuzzi:2024wyh}). 
    For temperature and $E$-mode polarization, we use \textit{Planck} at scales $2 \le \ell \le 50$ over $f_{\rm sky}=0.7$.
    At higher multipoles, we include two contributions.
    In the sky region observed by SO LAT ($f_{\rm sky}=0.4$), we adopt the combined SO LAT$+$\textit{Planck} noise curves provided by the SO collaboration\footnote{\url{https://github.com/simonsobs/so_noise_models}} for the baseline noise level~\cite{SimonsObservatory:2018koc}. These include the effect of extragalactic foreground removal, relevant at small angular scales where foregrounds dominate the signal, and are used in the multipole range $50 \le \ell \le 3000$.
    To account for the sky area not covered by SO LAT, we include an additional \textit{Planck}-only contribution with $f_{\rm sky}=0.3$, used at multipoles $50 \le \ell \le 2500$. The corresponding {\it Planck} noise power spectra for temperature and $E$-mode polarization are modeled using eq.~\eqref{eq:noise}. We consider the same frequency channels used in the SO analysis (see table~IV of ref.~\cite{Allison:2015qca} for the corresponding sensitivities and beam widths), and combine them via inverse-noise weighting to obtain the total noise curves. 
    Finally, for $B$-mode polarization we use the SO SATs noise curve for baseline sensitivity over the multipole range $30 \le \ell \le 300$, with $f_{\rm sky}=0.1$. We consider the conservative case in which no delensing procedure is applied to the $B$-mode spectra. A partial removal of the lensing $B$-mode contribution would further improve the sensitivity to primary $B$-modes~\cite{SimonsObservatory:2018koc,Smith:2010gu,LiteBIRD:2023aov}, thereby enhancing the constraining power on defect-induced signals.
    \item {\it LiteBIRD}: We construct the instrumental noise power spectra using the sensitivities and beam widths of the frequency channels between 78 and 195 GHz, as reported in table 13 of~\cite{LiteBIRD:2022cnt}. 
    For each channel, we assume $\sigma_T = \sigma_{E/B}/\sqrt{2}$ and compute the noise spectra according to eq.~\eqref{eq:noise}. 
    The individual channels are then combined through inverse-noise weighting to obtain effective temperature and polarization noise curves. 
    We adopt $f_{\rm sky}=0.7$ and include temperature and $E$-mode polarization in the multipole range $2 \le \ell \le 1000$, together with $B$-mode polarization for $2 \le \ell \le 150$ (see e.g.~\cite{LiteBIRD:2022cnt,LiteBIRD:2023zmo}).
\end{itemize}
Using the {\it Planck}$+$SO configuration, we assess the expected sensitivity to the cosmic string scenarios discussed in the previous sections.
In particular, we focus on NG and AH string networks, and investigate whether the inclusion of string loops in the NG case leads to a potentially observable imprint in future CMB data.
We do not expect the {\it Planck}$+$SO configuration to lead to a significant improvement in the constraints on domain wall networks.
Indeed, domain walls mainly source CMB anisotropies at the largest angular scales (see figure~\ref{fig:Dl_TT_stable_data}), where the sensitivity is dominated by \textit{Planck}.
Moreover, given the current upper bound on $A_{\rm DW}$, the corresponding contribution to the $B$-mode power spectrum is constrained to be extremely small at the angular scales probed by the SO SATs.
For this reason, forecasts for stable domain walls are instead presented for the \textit{LiteBIRD} satellite.

For our MCMC analysis with \texttt{Cobaya}, we use the mock CMB likelihood introduced in~\cite{Rashkovetskyi:2021rwg}.\footnote{\url{https://github.com/misharash/cobaya_mock_cmb}} Simulated data are constructed as the sum of fiducial theoretical spectra and the experimental noise contributions:
\begin{equation}
    \widehat{C}_\ell^{\,XX} = C_\ell^{XX, \, \mathrm{fid}} + N_\ell^{XX} \, .
    \label{eq:mock_data}
\end{equation}
As our fiducial cosmology, we adopt the mean values of the $\Lambda$CDM parameters obtained from our baseline analysis with \textit{Planck}$+$BK18 (see table~\ref{tab:constraints}), setting both the tensor-to-scalar ratio and the string/wall tension to zero ($r = 0$, $A_{\rm defect} = 0$).

\subsection{Results and discussion}
\label{sec:forecasts_results}
We now present the forecasted constraints for the experimental configurations described in section~\ref{sec:forecasts_noise}.
The resulting 95\% credible upper limits on $A_{\rm str}$ and $A_{\rm DW}$, together with the corresponding limits on $G\mu$ and $\sigma^{1/3}$, are summarized in table~\ref{tab:forecasts_defects}.
\begin{table}[!t]
    \centering
    \resizebox{0.85\textwidth}{!}{
    \begin{tabular}{ c c c c c }
        \hline
        \hline
        \noalign{\vskip 1mm}
        & \multicolumn{3}{c}{\textbf{\small \textit{Planck}$+$SO}}
        & \textbf{\small \textit{LiteBIRD}} \\
        \cmidrule(lr){2-4}\cmidrule(lr){5-5}
        \noalign{\vskip 1mm}
        Parameter
        & NG strings
        & NG strings with loops
        & AH strings
        & Domain walls \\
        \noalign{\vskip 1mm}
        \hline
        \noalign{\vskip 1mm}
        $A_{\rm str}$ & $< 1.8 \times 10^{-3}$ & $< 1.4 \times 10^{-3}$ & $< 5.4 \times 10^{-3}$ & -- \\[1mm]
        $G\mu$        & $< 4.2 \times 10^{-8}$ & $< 3.8 \times 10^{-8}$ & $< 7.3 \times 10^{-8}$ & -- \\[1mm]
        \noalign{\vskip 1mm}
        \hdashline
        \noalign{\vskip 1.5mm}
        $A_{\rm DW}$  & -- & -- & -- & $< 0.16$ \\[1mm]
        $\sigma^{1/3}\;[\mathrm{MeV}]$ & -- & -- & -- & $< 0.74$ \\[1mm]
        \hline
        \hline
    \end{tabular}
    }
    \caption{Forecasted 95\% credible upper limits on $A_{\rm str}$ and $A_{\rm DW}$, together with the corresponding derived limits on the string tension $G\mu$ and the domain wall tension $\sigma^{1/3}$, for the \textit{Planck}$+$SO and \textit{LiteBIRD} configurations described in section~\ref{sec:forecasts_noise}.}
    \label{tab:forecasts_defects}
\end{table}

\begin{figure}[!t]
    \centering
    \includegraphics[width=0.5\linewidth]{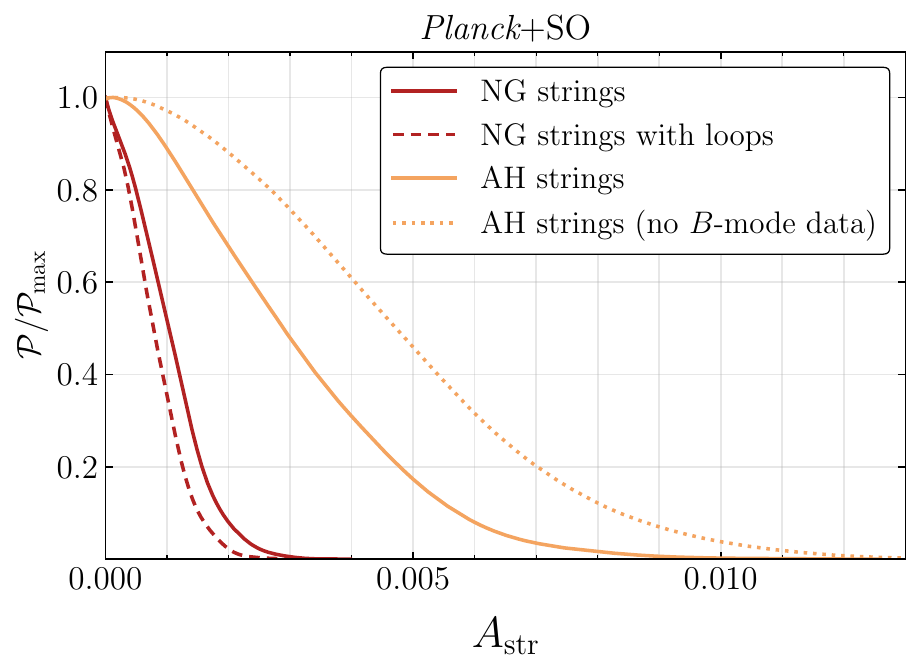}
    \caption{One-dimensional posterior distributions of $A_{\rm str}$ from our {\it Planck}$+$SO forecast analysis, for Nambu-Goto strings with and without loops (solid and dashed red lines, respectively), and for Abelian-Higgs strings (orange). The dotted orange line shows the $A_{\rm str}$ posterior for AH strings obtained without including $B$-mode data from the SO SATs.}
    \label{fig:triangle_forecasts_1D}
\end{figure}

We first discuss the results for cosmic string networks from \textit{Planck}$+$SO.
Figure~\ref{fig:triangle_forecasts_1D} shows the marginalized one-dimensional posterior distributions for $A_{\rm str}$ for NG strings (with and without loops) and AH strings. The full triangle plot and the corresponding constraints on all cosmological parameters sampled in our analysis are reported in appendix~\ref{app:SO}.
For NG strings, we find that the constraint on $A_{\rm str}$ strengthens by nearly an order of magnitude with respect to \textit{Planck}$+$BK18 data, corresponding to an improvement slightly larger than a factor of three in the bound on the string tension, $G\mu$.
Including string loops further tightens this limit by about 9.5\%, in contrast to what is observed with current \textit{Planck}$+$BK18 data, where the impact of loops is negligible. This is driven by the enhanced sensitivity of SO LAT at small angular scales, where the contribution from string loops becomes relevant.
For AH strings, we find that the upper bound on $G\mu$ improves by a factor of about 2.5.

We also assess the impact of $B$-mode polarization from the SO SATs.
When both $A_{\rm str}$ and $r$ are varied, the inclusion of $B$-mode data strengthens the constraint on $A_{\rm str}$ for AH strings by a factor of 1.4 with respect to the case in which $B$-modes are not included. 

\begin{figure}[!t]
    \centering
    \includegraphics[width=0.5\linewidth]{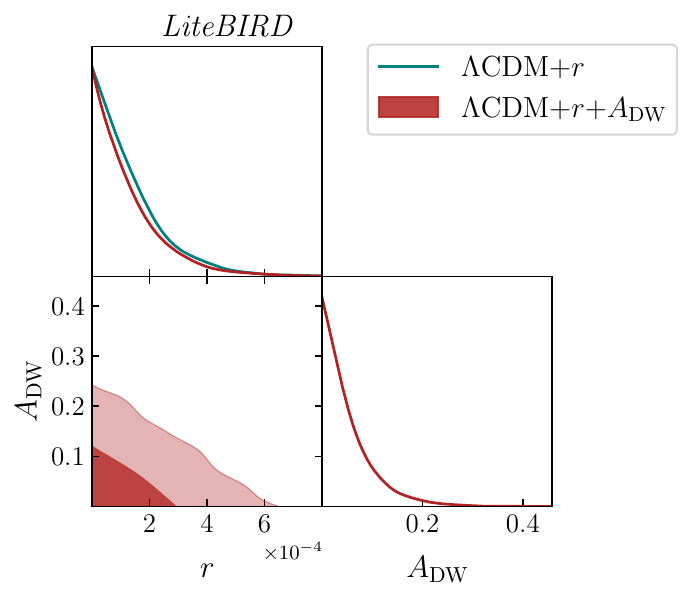}
    \caption{One- and two-dimensional posterior distributions for $A_{\rm DW}$ and the tensor-to-scalar ratio $r$ from the \textit{LiteBIRD} forecast analysis for stable domain wall networks.}
    \label{fig:DWs_Litebird_A_DW_r}
\end{figure}

We now turn to stable domain wall networks and present the forecasts obtained with \textit{LiteBIRD}.
Figure~\ref{fig:DWs_Litebird_A_DW_r} shows the one- and two-dimensional posterior distributions for $A_{\rm DW}$ and the tensor-to-scalar ratio $r$, while the full triangle plot and the corresponding parameter constraints are reported in appendix~\ref{app:SO}.
From this analysis, we obtain a 95\% credible upper limit of $A_{\rm DW} < 0.16$, which translates into a bound on the domain wall tension of $\sigma^{1/3} < 0.74 \; \mathrm{MeV}$. 
This represents an improvement by a about a factor of five on the domain wall tension $\sigma$, compared to the current \textit{Planck}$+$BK18 bound. 
A weak anti-correlation between $A_{\rm DW}$ and $r$ is visible in figure~\ref{fig:DWs_Litebird_A_DW_r}.
As a result, allowing for a non-zero domain wall contribution leads to a slightly tighter upper bound on $r$ relative to the baseline $\Lambda$CDM$+r$ case (see table~\ref{tab:forecasts_Litebird}).

\section{Implications for Gravitational Wave Observatories}
\label{sec:Implications for other non-CMB GW observations}

In this section, we briefly discuss the implications of our CMB results for gravitational wave observations. 
A NG string network with a tension as high as the best-fit value from \textit{Planck}$+$BK18, $G\mu=7.8\times 10^{-8}$, would generate a Stochastic Gravitational Wave Background (SGWB) with an amplitude in the $n$Hz band that would exceed that of the signal recently reported by pulsar timing experiments~\cite{NANOGrav:2023gor,Reardon:2023gzh,EPTA:2023fyk,Xu:2023wog}. The analysis presented in~\cite{EPTA:2023xxk,NANOGrav:2023hvm} shows that the observed signal would be compatible with the SGWB generated by the loops formed in the evolution of a NG string network with $G\mu \sim \mathcal{O}(10^{-10})$, assuming that they decay by emitting only gravitational waves. Indeed, larger values of string tension could be reconciled with gravitational wave data if one considers the possibility that string loops may also decay via the emission of non-gravitational radiation and that, as a result, only a fraction $f_{\rm NG}<1$ of the loops contributes to the SGWB (as suggested in~\cite{Hindmarsh:2021mnl} for AH strings). An analysis of this scenario was performed in~\cite{Kume:2024adn}, where it was found that, besides the previously mentioned case of a network with $f_{\rm NG}=1$ and $G\mu \sim \mathcal{O}(10^{-10})$, NANOGrav data could also be explained by the SGWB generated from a network with $f_{\rm NG}\simeq 0.03$ and $G\mu \sim \mathcal{O}(10^{-6})$, a value of the tension that however exceeds our 95\% credible upper bound by about one order of magnitude. 

\begin{figure}[!t]
    \centering
    \includegraphics[width=0.7\linewidth]{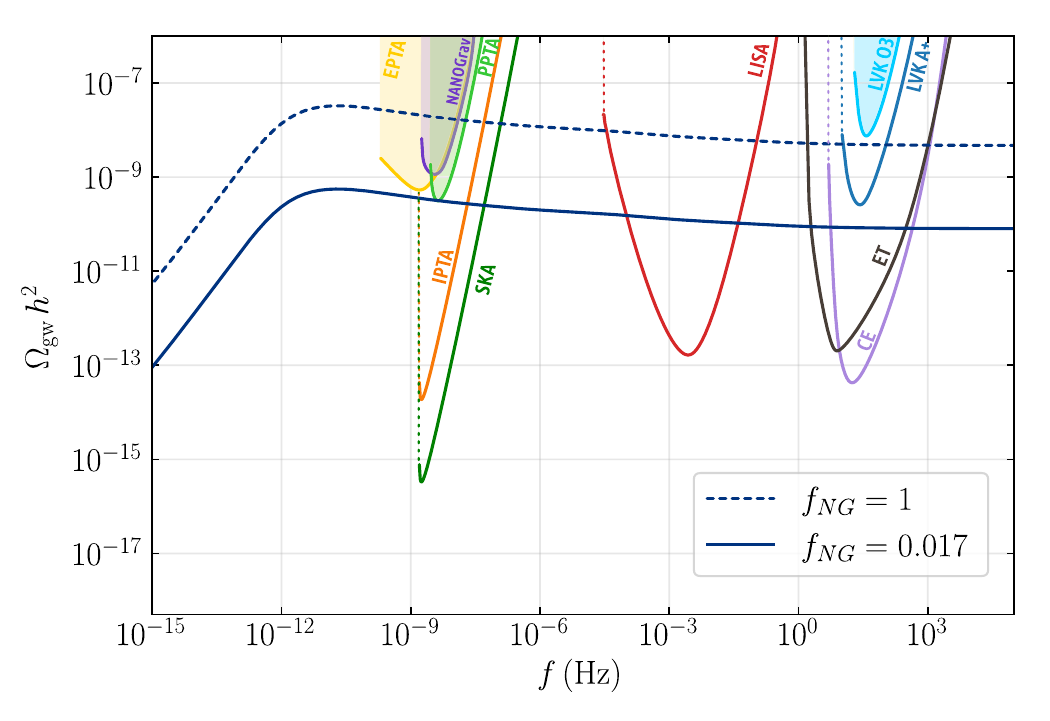}
    \caption{Stochastic gravitational wave background generated by a Nambu-Goto string network with a tension $G\mu=7.8\times10^{-8}$ (best-fit value obtained from the \textit{Planck}$+$BK18 analysis) with $f_{\rm NG}=1$ (dashed dark-blue line) and $f_{\rm NG}=0.017$ (solid dark-blue line), alongside the power-law integrated sensitivity windows of current and upcoming gravitational wave detectors~\cite{LISA:2017pwj,Schmitz:2020syl,KAGRA:2021kbb}. The shaded yellow, purple, green and light-blue areas represent, respectively, the sensitivity windows of the European Pulsar Timing Array (EPTA)~\cite{EPTA:2023fyk,EPTA:2023xxk}, the North American Nanohertz Observatory for Gravitational Waves (NANOGrav)~\cite{NANOGrav:2023ctt}, Parkes Pulsar Timing Array (PPTA)~\cite{Manchester:2012za} and the LIGO-Virgo-Kagra (LVK) O3 run~\cite{KAGRA:2021kbb}. The medium-blue, orange, dark-green, black, red and light-purple lines represent, respectively, the forecasted sensitivity curves of LVK's $A^+$ design~\cite{KAGRA:2021kbb}, the International Pulsar Timing Array (IPTA)~\cite{Hobbs:2009yy,Verbiest:2016vem}, the Square Kilometer Array (SKA)~\cite{Janssen:2014dka,Weltman:2018zrl}, Einstein Telescope (ET)~\cite{Punturo:2010zz,ET:2019dnz}, the Laser Interferometer Space Antenna (LISA)~\cite{LISA:2017pwj} and Cosmic Explorer (CE)~\cite{Reitze:2019iox}.}
    \label{fig:GWs}
\end{figure}

The only remaining option is that the SGWB from this network has an amplitude below that probed by current pulsar timing arrays. Using the semi-analytical model to compute the cosmic string SGWB in~\cite{Sousa:2013aaa,Sousa:2020sxs} (whose parameters may be calibrated to describe NG networks\footnote{This corresponds to setting in~\cite{Sousa:2013aaa,Sousa:2020sxs} the loop-size parameter to $\alpha=0.34$, the \textit{fuzziness} parameter to $\mathcal{F}=0.1$ and fixing the spectral index to $q=4/3$. In our computation, we also included the impact of the change in relativistic degrees of freedom as the universe expands, assuming the particle contents of the standard model.}), we find that, in order to avoid detection in this frequency band for a tension value corresponding to the \textit{Planck}$+$BK18 best-fit, one would need roughly $f_{\rm NG}<0.017$. 
We plot the SGWB energy density spectrum, defined as $\Omega_{\rm gw}(f) \equiv (1/\rho_c) (d\rho_{\rm gw}/d\ln f)$, alongside the sensitivity windows of current and upcoming GW detectors~\cite{LISA:2017pwj,Schmitz:2020syl,KAGRA:2021kbb,NANOGrav:2023ctt}\footnote{For visualization purposes, we did not include the Big Bang Observer~\cite{Harry:2006fi} and Deci-hertz Interferometer Gravitational wave Observatory~\cite{Kawamura:2006up} in this plot, but we briefly discuss them in the text.} in figure~\ref{fig:GWs}. Interestingly, such a network would also avoid detection in the LIGO-Virgo-Kagra O3 run and would be below their forecasted $A^+$ design power-law integrated sensitivity window~\cite{KAGRA:2021kbb}, but could potentially be detected in the near future by upcoming detectors. In particular, we find that it would be within the reach of the Laser Interferometer Space Antenna, the International Pulsar Timing Array, Einstein Telescope and Cosmic Explorer provided that $f_{\rm NG}>10^{-5}-10^{-6}$. The Square Kilometer Array, Big Bang Observer and the Deci-hertz Interferometer Gravitational wave Observatory could probe this spectrum even further, up to $f_{\rm NG}\sim10^{-8}-10^{-9}$. Such a detection of GWs, or of CMB (or other) signatures of a string network with $G\mu \gtrsim \mathcal{O}(10^{-10})$, consistent with the tension value preferred by our \textit{Planck}$+$BK18 analysis, would provide strong evidence that the role of non-gravitational radiation may be significant and, therefore, that the NG approximation may fail to capture the physics of realistic cosmic string networks. 

As to stable domain walls, the constraints derived here imply that one should not expect a detection with other, non-CMB, gravitational wave experiments in the near future: they imply that their abundance today, $\Omega_{\rm DW}^0$, should be smaller than $10^{-4}$ and, as a result, that their SGWB should have a very small amplitude. Moreover, these walls are expected to generate a GW background that peaks at frequencies $\mathcal{O}(10^{-18}) \, \rm{Hz}$~\cite{Ferreira:2023jbu,Gruber:2024pqh} --- accessible only to CMB experiments --- and that decreases quickly with increasing frequency, thus generating a negligible signal in the frequency ranges probed by current and planned gravitational wave experiments.\footnote{Note that this is not the case for biased domain walls that fully decay before the present time, as these would avoid our stringent constraints and contribute dominantly at higher frequencies (see e.g.~\cite{ZambujalFerreira:2021cte,Ferreira:2022zzo,Ferreira:2023jbu,Ferreira:2024eru,Notari:2025kqq,Blasi:2025tmn}).}

\section{Conclusions}
\label{sec:conclusions}

In this work, we have presented an updated and comprehensive analysis of CMB constraints on stable networks of cosmic strings and domain walls, combining the full \textit{Planck} 2018 temperature and polarization data~\cite{Planck:2019nip} with BICEP/\textit{Keck} 2018 (BK18) $B$-mode measurements~\cite{BICEP:2021xfz}, and extending the analysis with forecasts for the upcoming Simons Observatory (SO) experiment~\cite{SimonsObservatory:2018koc}. To the best of our knowledge, this is the first time that these datasets have been jointly used to probe cosmic defect scenarios.

As a result, we have derived 95\% credible upper limits of $G\mu < 1.2 \times 10^{-7}$ for Nambu-Goto strings, $G\mu < 1.8 \times 10^{-7}$ for Abelian-Higgs strings, and $\sigma^{1/3} < 1.3 \; \mathrm{MeV}$ for stable domain wall networks. These bounds correspond to the conservative choice of $L_f = 0.5$, for the parameter that controls the decay rate of defects in the Unconnected Segment Model. We have also explored the case $L_f = 1$ (for which the CMB anisotropies are maximal), finding systematically tighter constraints, consistent with previous studies.
Overall, our results improve previous CMB constraints on the string power spectra by up to 30\%, and correct previous analyses of domain wall networks that overestimated the domain wall power spectrum by a factor of about ten at large scales, and larger towards smaller scales.
A key new ingredient in our analysis is the inclusion, for the first time, of $B$-mode polarization measurements from BK18.
The impact of $B$-modes is particularly significant for cosmic strings, especially in the case of Abelian-Higgs strings, for which the inclusion of BK18 data strengthens the constraints on the string tension by about $18\%$ compared to analyses based solely on temperature and $E$-mode polarization from \textit{Planck}. In light of the increasing importance of $B$-mode polarization data, it would be valuable to re-analyse the case where unequal-time correlators (UETC) are directly extracted from field-theory simulations and where one would expect comparable improvements in constraining power. 

Notably, we find a mild preference for models with a non-zero cosmic string tension over the baseline $\Lambda$CDM (or $\Lambda$CDM$+r$) scenario.
The corresponding best-fit values lie in the range $G\mu \sim (8\text{--}11)\times 10^{-8}$, depending on the dataset combination and the string model. 
This mild preference, driven by \textit{Planck} data, appears for both Abelian-Higgs and Nambu-Goto strings, despite the substantial differences in their predicted CMB power spectra. 
Although the statistical significance of this preference is low, it will be interesting to assess whether it persists in future analyses with improved data or with more refined modeling. 

Indeed, a key assumption of this work is the modeling of Nambu-Goto strings, Abelian-Higgs strings, and stable domain wall networks within the Unconnected Segment Model, combined with a description of their large-scale evolution through the corresponding simulation-calibrated Velocity-dependent One-Scale (VOS) model. This framework has been widely used in previous CMB analyses and has been shown to yield results in good agreement with those obtained using UETCs extracted directly from field-theory simulations. It also has the advantage that the impact of the radiation-matter and matter-$\Lambda$ transitions --- which was shown in~\cite{Lizarraga:2016onn} to be vital to explore the full constraining power of CMB data --- is naturally taken into account, while the use of UETCs directly extracted from simulations often requires interpolating between the results of different simulations and/or performing extrapolations due to limitations in dynamical range. 

It should also be noted that while Nambu-Goto strings provide an efficient phenomenological description of thin local defects, Abelian-Higgs simulations include additional radiative channels and therefore offer a richer description of the underlying microphysics. As a result, the corresponding constraints can be regarded as a more appropriate proxy for other radiative models such as global string scenarios, which are known to produce CMB power spectra with shapes that share similarities to those of Abelian-Higgs strings~\cite{Lopez-Eiguren:2017dmc}. This motivates a dedicated analysis of global strings, in particular, in view of the current uncertainty on the evolution of the effective string tension and string density~\cite{Hindmarsh:2021vih,Saikawa:2024bta,Gorghetto:2024vnp}. 
Note however that since the expressions for the stress-energy tensor in section~\ref{sec:USM} apply only to strings whose energy is localized at the core, at present, one cannot apply this methodology to the case of global strings and should instead resort to the determination of the UETC from numerical simulations.

Besides the analysis with current data, we have also presented the first dedicated forecasts for cosmic defect searches with upcoming CMB experiments. 
For cosmic strings, we have considered the {\it Planck}$+$SO configuration~\cite{SimonsObservatory:2018koc}, including $B$-mode measurements from the SO SATs.
We have found that the bounds on the string tension could improve by slightly more than a factor of three for Nambu-Goto strings and by about $2.5$ for Abelian-Higgs strings.
Moreover, the enhanced sensitivity of the SO LAT at small angular scales allows the contribution from string loops to become relevant, leading to a further tightening of the constraints at the $\sim 10\%$ level.
In this respect, a natural extension of the present analysis is to include high-resolution measurements of CMB anisotropies from current ground-based experiments, such as the Atacama Cosmology Telescope (ACT)~\cite{AtacamaCosmologyTelescope:2025blo} and the South Pole Telescope (SPT)~\cite{SPT-3G:2025bzu}. These provide increased constraining power at small angular scales and can be particularly relevant for testing cosmic string scenarios.

For stable domain wall networks, we have presented forecasts for the {\it LiteBIRD} satellite~\cite{LiteBIRD:2022cnt}. In this case, the improved sensitivity to large-scale polarization strengthens the bound on the wall tension, $\sigma$, by about a factor of five compared to the current \textit{Planck}$+$BK18 limit.

Finally, we also discussed the evidence for GWs in the Pulsar Timing Array (PTA) data~\cite{EPTA:2023xxk,NANOGrav:2023hvm}, which provides constraints on the string tension $G \mu <10^{-10}$ that are stronger than those derived in this work. We noted that these constraints are applicable for string networks where all loops are long-lived and contribute to the stochastic gravitational wave background. 
If string loops can decay via the emission of
non-gravitational radiation and only a fraction contributes to the stochastic gravitational wave background, as in the Abelian-Higgs case~\cite{Kume:2024adn}, a loop fraction of 
$f_{\rm NG}<0.017$ would be enough for the \textit{Planck}$+$BK18 best-fit string tension to avoid detection at PTAs~\cite{Sousa:2013aaa,Sousa:2020sxs}.

Overall, our results reinforce the role of CMB $B$-mode polarization as a powerful probe of topological defects, and provide a consistent and complete methodology for future analysis.

\vspace{0.4cm}
\noindent{\bf Note added:}
While this paper was being finalized, a related analysis appeared in~\cite{Raidal:2026cpb}, which also derives CMB constraints on Nambu-Goto strings and considers the combination of \textit{Planck} and ACT DR6 data.
There, unequal-time correlators are computed analytically within the Unconnected Segment Model and provided as input to \texttt{CAMB}. This approach requires assuming $L_f = 1$, a choice known to yield tighter bounds on the string tension (as we also show). They also treat the chopping factor $\cc$ and the wiggliness parameter $\alpha$,  defined respectively in eqs.~\eqref{eq:VOSL} and~\eqref{eq:alpja},  as free parameters in their analysis and, as a result, these constraints cannot be directly compared. However, when a linear flat prior for $G\mu$ is adopted in their analysis, with Gaussian priors for $\cc$ and $\alpha$ around the values $\cc=0.23$ (coinciding with ours) and $\alpha=1.7$,  one may see, using eq.~\eqref{eq:alpja}, that the results of both analyses with \textit{Planck}-only data are consistent with each other.

\acknowledgments

L.C. thanks Margherita Lembo for useful discussions. L.S. thanks Pedro Avelino, Tasos Avgoustidis, José Correia and Ivan Rybak for enlightening discussions. C.W. also thanks Pedro Avelino for helpful discussions. L.C. and R.Z.F acknowledge the financial support provided by FCT - Fundação para a Ciência e Tecnologia (FCT), I.P., through  
the Strategic Funding UID/04650/2025 and UID/04564/2025 and 
national funds with DOI identifiers 10.54499/2023.11681.PEX, 10.54499/2024.00249.CERN funded by measure RE-C06-i06.m02 -- ``Reinforcement of funding for International Partnerships in Science, Technology and Innovation'' of the Recovery and Resilience Plan -- RRP, within the framework of the financing contract signed between the Recover Portugal Mission Structure (EMRP) and the Foundation for Science and Technology I.P. (FCT), as an intermediate beneficiary, as well as the advanced computing projects 2024.00249.CERN.F1 and 2024.07037.CPCA.A1. C.W. is supported by FCT (\url{https://ror.org/00snfqn58}) through the PhD fellowship with reference UI/BD/154758/2023 (\url{https://doi.org/10.54499/UI/BD/154758/2023}). L.S. and C.W. are also funded by FCT through the Strategic Funding UID/04434/2025 and the research grant 2024.17828.PEX - \textit{Unveiling the early universe with topological defects} (\url{https://doi.org/10.54499/2024.17828.PEX}). For the purpose of Open Access, the authors have applied a CC-BY public copyright license to any Author's Accepted Manuscript (AAM) version arising from this submission.


\appendix

\section{Including cosmic string loops in the USM}
\label{app:loops}
To include the impact of cosmic string loops on the anisotropies generated from NG string networks, we will follow the approach introduced in~\cite{Rybak:2021scp}. Therein it is assumed that decaying segments, instead of quickly disappearing as in the original USM model, are converted into closed loops of string that slowly decay by emitting gravitational radiation, seeding perturbations actively for a longer period. NG numerical simulations~\cite{Blanco-Pillado:2013qja} indeed indicate that $10\%$ of the energy lost by the network goes into long-lived loops of string --- created with a comoving length of $l_c^b\simeq 0.34 L_c$ --- and the remainder goes into much smaller loops. Since such small loops are expected to have a negligible impact on CMB anisotropies~\cite{Rybak:2021scp}, here we will consider only the potential contribution of the population of large loops. 

We will assume that, at each timestep $\tau_i$, a fraction of $0.1$ of the total length of the decaying string segments is converted into circular loops of strings with a comoving radius of $R_c(\tau_i)=l_c^b(\tau_i)/(2\pi)$. The number of such loops created at each step is then given by~\cite{Rybak:2021scp}:
\be
\mathcal{N}_l (\tau_i)=0.1 \frac{\mathcal{N}(\tau_i) L_c(\tau_i)}{l_c^b(\tau_i)}=\frac{\mathcal{N}(\tau_i)}{3.4}\,,
\ee 
where $\mathcal{N}(\tau_i)$ is the number of decaying string segments in~\eqref{eq:Ni}. Cosmic string loops are generally expected to emit a roughly constant power in gravitational radiation, given by $P=\Gamma G\mu^2$, with $\Gamma=50$~\cite{Quashnock:1990wv,Casper:1995ub}. So, the radius of the loops is expected to progressively decrease as
\be 
R_c(\tau)=\frac{1}{2\pi a(\tau)}\left[0.34L_c(\tau_i)a(\tau_i)-\Gamma G\mu (t(\tau)-t(\tau_i))\right]\,,
\label{eq:Rcevo}
\ee 
until they eventually evaporate at a time $\tau_f$ determined by $R_c(\tau_f)=0$. Therefore, we need to take into account their contribution to the CMB anisotropies for $\tau_i\le\tau\le\tau_f$.

The creation and evaporation of cosmic string loops is, in this framework, implemented very similarly to segment decay in the original USM. In fact, the total stress-energy tensor of loops is written as:
\begin{equation}
    \label{eq:USMloop}
    \Theta_{\mu\nu}^{\rm loops}(\mathbf{k},\tau) = \sum_{i=1}^{N}\sqrt{\mathcal{N}_l(\tau_i)} \tilde{\Theta}_{\mu\nu}^i(\mathbf{k},\tau) T^{\rm off}(\tau,\tau_i,L_f) T^{\rm on}(\tau,\tau_i,L_i)\, ,
\end{equation}
where $T^{\rm off}(\tau,\tau_i,L_f)$ assumes the form in~\eqref{eq:Toff} and function $T^{\rm on}(\tau,\tau_i,L_i)$ is a function that ``turns on'' the contribution of the loops created at a time $\tau_i$, given by
\begin{align}
    T^{\rm on}(\tau,\tau_i,L_i) = 
        \begin{cases}
            0 &\tau < \tau_i \\
            \frac{1}{2} + \frac{1}{4}(3z-z^3) \quad & \tau_i \le \tau < L_i\tau_i \\
            1 & L_i \tau_i \le \tau
        \end{cases} \, ,
\end{align}
with
\be 
z=\frac{2\ln(\tau_i/\tau)}{\ln(1/L_i)}-1\,.
\ee 
Here, $L_i$ is a constant that determines how fast loops appear and varies from $0$ at $\tau_i$ to $1$ at $L_i\tau_i$. To ensure that loops appear at the same rate string segments disappear, we should have that $L_i=2-L_f$. Moreover, as explained in detail in~\cite{Rybak:2021scp}, all the loops that are created at a given timestep are also consolidated into a single loop. Thus, the summation in~\eqref{eq:USMloop} is also performed over timesteps and $ \tilde{\Theta}_{\mu\nu}^i(\mathbf{k},\tau)$ is the stress-energy tensor of the consolidated loop created at $\tau_i$.

To compute $ \tilde{\Theta}_{\mu\nu}^i(\mathbf{k},\tau)$, let us assume that the loops are circular and planar and that their translational velocity is perpendicular to them. We may then define a set of orthogonal unitary vectors, $(\un{u},\un{w},\un{v})$, such that the loop is contained in the plane generated by $\un{u}$ and $\un{w}$ and $\un{v}$ coincides with the direction of the velocity. In this case, the components of the stress-energy tensor of the loop may, in Fourier space, be written as~\cite{Rybak:2021scp}
\bq 
\tilde{\Theta}_{00} (\mathbf{k},\tau) & = &  2\pi \mu R_c \gamma_l J_0 (\chi) \cos \left(\mathbf{k}\cdot\xx_0^l + v_l k \tau \uni{v}{3} \right)\,,\label{eq:Theta00loop}\\
\tilde{\Theta}_{ij} (\mathbf{k},\tau) & = & \tilde{\Theta}_{00} (\mathbf{k},\tau) \left[v_l^2 \uni{v}{i}\uni{v}{j} -\frac{1}{2\gamma_l^2} \left(\uni{u}{i}\uni{u}{j}\mathcal{I}_- + \uni{w}{i}\uni{w}{j}\mathcal{I}_++2\mathcal{I}\uni{u}{(i}\uni{w}{j)}\right)\right]\label{eq:Thetaijloop}\,,
\eq 
where $R_c$ is the comoving radius of the loop, $v_l$ is the center-of-mass velocity of the loop, $\gamma_l=(1-v_l^2)^{-1/2}$ and $\xx_0^l$ is the position of the center-of-mass of the loop. Moreover, we have defined
\be
\mathcal{I}_{\pm}=1\pm \frac{J_2(\chi)}{J_0(\chi)} \cos{(2B)}\,, \quad \mathcal{I}=\frac{J_2(\chi)}{J_0(\chi)} \sin{(2B)}\,,\quad\mbox{and} \quad \uni{u}{(i}\uni{w}{j)}=\frac{1}{2}\left(\uni{u}{i}\uni{w}{j}+\uni{w}{i}\uni{u}{j}\right)
\ee
where
\be 
\chi=k R_c \left(\uni{u}{3}^2+\uni{w}{3}^2 \right)^{1/2}\,,\quad B=\arctan{\left(\frac{\uni{u}{3}}{\uni{w}{3}}\right)}\,,
\ee 
and $J_n(\cdots)$ is the Bessel function of the first kind.

Numerical simulations indicate that loops are created with a translational velocity of $v_l(\tau_i)\simeq 0.42$~\cite{Blanco-Pillado:2011egf}. However, after loops are created, this velocity should decrease as $v_l\gamma_l\propto a^{-1}$ as a result of the expansion of the background. So, when implementing loops in the USM, we assume that the loop velocity evolves according to
\be 
v_l(\tau)=v_l(\tau_i)\left\{v_l^2(\tau_i)+\left[1-v_l^2(\tau_i)\right]\left(\frac{a(\tau)}{a(\tau_i)}\right)^2\right\}^{-1/2}\,,
\ee 
while the evolution of $R_c$ is given by~\eqref{eq:Rcevo}. Moreover, as shown in~\cite{Wu:1998mr}, the positions and velocities of loops are highly correlated with those of long strings and including these correlations is vital to determine the CMB anisotropies accurately~\cite{Rybak:2021scp}. To account for the fact that loops are created along the long strings, we will assume that loops are created at the same position where the segments decayed. In other words, we assume that the center of mass of the consolidated loop coincides with the center of mass of the decaying consolidated segment at the time of decay: $\xx_o^l=\xx_0+vk\tau\hat{\xd}$. Moreover, since loops tend to move in the same direction as the string from which they have been chopped off, we assume that the consolidated loop ``inherits'' the direction of the velocity of the decaying segment, such that $\un{v}=\hat{\xd}$ (while $\un{u}$ and $\un{w}$ are randomly attributed).

The total stress-energy tensor of a NG string network including the contribution of loops --- i.e., the source used when running \texttt{CMBACT4} in this case --- is given by: 
\be 
\Theta_{\mu\nu}^{\rm total}(\mathbf{k},\tau)=\Theta_{\mu\nu}^{\rm network}(\mathbf{k},\tau)+\Theta_{\mu\nu}^{\rm loops}(\mathbf{k},\tau)\,,
\ee 
where $\Theta_{\mu\nu}^{\rm network}(\mathbf{k},\tau)$ is given by~\eqref{eq:USM} and $\Theta_{\mu\nu}^{\rm loops}(\mathbf{k},\tau)$ is computed as described in this appendix. Note that we do not include loops in the computation of the anisotropies generated by the AH network because simulations indicate that these loops are short-lived~\cite{Hindmarsh:2021mnl,Blanco-Pillado:2023sap,Baeza-Ballesteros:2024sny} and thus should not give a significant contribution. The impact of loops on the CMB anisotropies is shown in figures \ref{fig:Strings_NG_loops_1D} and \ref{fig:triangle_forecasts} and in tables \ref{tab:forecasts_defects} and  \ref{tab:forecasts}.

\section{Triangle plots for current constraints with $L_f=0.5$ and $L_f=1$}
\label{appendix:triangle_plots}
For completeness, this appendix collects the full triangle plots for all defect models and dataset combinations discussed in the main text, for the case with $L_f = 0.5$. We also include the corresponding results for $L_f = 1$ for the \textit{Planck}$+$BK18 configuration with $r$ left free to vary, shown as dashed green lines and contours in each plot.
The full set of cosmological and defect parameter constraints for the \textit{Planck}$+$BK18 case with $r$ free are reported in table~\ref{tab:constraints}, while table~\ref{tab:constraints_Lf1} lists the corresponding limits on the defect amplitudes and tensions for $L_f = 1$.

\begin{figure}[!t]
    \centering
    \includegraphics[width=0.45\linewidth]{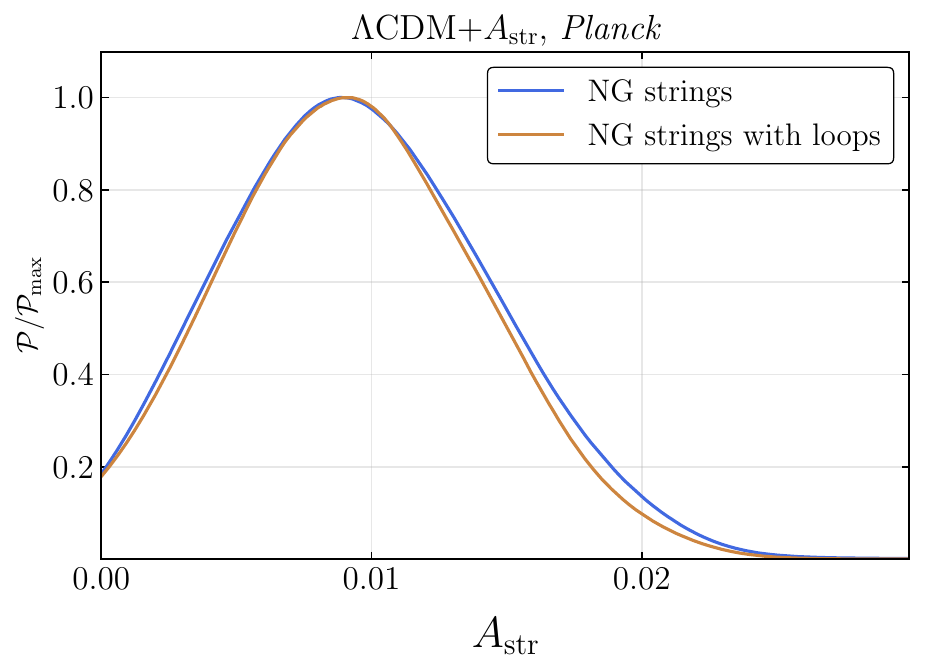}
    \caption{One-dimensional posterior of $A_{\rm str}$ for NG strings with and without the inclusion of loops in the USM, using {\it Planck} data.}
    \label{fig:Strings_NG_loops_1D}
\end{figure}

Figure~\ref{fig:Strings_NG_loops_1D} shows the comparison between the one-dimensional posterior distributions of $A_{\rm str}$ for NG strings, with and without the inclusion of string loops in the USM (as discussed in appendix~\ref{app:loops}), obtained with {\it Planck} data.

\begin{table}[!t]
    \centering
    \textbf{\small \textit{Planck}$+$BK18} \\[0.75mm]
    \resizebox{\textwidth}{!}{
    \begin{tabular}{ c c c c c } 
        \hline
        \hline
        \noalign{\vskip 1mm}
        Parameter & $\Lambda$CDM$+r$ & $+$Domain walls & $+$NG strings & $+$AH strings \\
        \noalign{\vskip 1mm}
        \hline
        \noalign{\vskip 1mm}
        $\omega_b$ & $0.02240 \pm 0.00014$ & $0.02240 \pm 0.00014$ & $0.02242 \pm 0.00014$ & $0.02242 \pm 0.00014$ \\[1mm]
        $\omega_c$ & $0.1196 \pm 0.0011$ & $0.1196 \pm 0.0011$ & $0.1194 \pm 0.0011$ & $0.1194 \pm 0.0011$ \\[1mm]
        $H_0 \; [\mathrm{km} \, \mathrm{s}^{-1} \, \mathrm{Mpc}^{-1}]$ & $67.51 \pm 0.50$ & $67.51 \pm 0.50$ & $67.61 \pm 0.51$ & $67.61 \pm 0.52$ \\[1mm]
        $\tau_{\rm reio}$ & $0.0582^{+0.0052}_{-0.0063}$ & $0.0581^{+0.0054}_{-0.0062}$ & $0.0580^{+0.0053}_{-0.0063}$ & $0.0581^{+0.0053}_{-0.0064}$ \\[1mm]
        $\ln(10^{10} A_s)$ & $3.051 ^{+0.011}_{-0.012}$ & $3.051^{+0.011}_{-0.012}$ & $3.045 \pm 0.012$ & $3.046 \pm 0.012$ \\[1mm]
        $n_s$ & $0.9665 \pm 0.0040$ & $0.9667 \pm 0.0039$ & $0.9655 \pm 0.0040$ & $0.9667 \pm 0.0039$ \\[1mm] 
        $r$ & $< 0.035$ & $< 0.035$ & $< 0.031$ & $< 0.030$ \\[1mm]
        \hline
        \noalign{\vskip 1mm} 
        $A_{\rm DW}$ & -- & $< 5.2$ & -- & -- \\[1mm]
        $\sigma^{1/3} \; [\mathrm{MeV}]$ & -- & $< 1.3$ & -- & -- \\[1mm]
        \noalign{\vskip 1mm}
        \hdashline
        \noalign{\vskip 1.5mm}
        $A_{\rm str}$ & -- & -- & $< 0.014$ & $< 0.031$ \\ [1mm]
        & & & ($6.04 \times 10^{-3}$) & ($1.19 \times 10^{-2}$) \\[1mm]
        $G\mu$ & -- & -- & $< 1.2 \times 10^{-7}$ & $< 1.8 \times 10^{-7}$\\[1mm]
        & & & ($7.8 \times 10^{-8}$) & ($1.1 \times 10^{-7}$) 
        \\[1mm]
        \hline
        \hline
    \end{tabular}
    }
    \caption{Constraints on the cosmological and defect parameters derived from \textit{Planck}$+$BK18 data. $\Lambda$CDM parameters are quoted as 68\% credible intervals, while 95\% credible upper limits are reported for $r$ and for the domain wall or string tension. Values between parentheses represent the best-fits for $A_{\rm str}$ and $G\mu$.}
    \label{tab:constraints}
\end{table}

\begin{table}[!t]
    \centering
    \textbf{\footnotesize Constraints for $L_f=1$, from \textit{Planck}$+$BK18} \\[0.75mm]
    \resizebox{0.6\textwidth}{!}{
    \begin{tabular}{ c c c c } 
        \hline
        \hline
        \noalign{\vskip 1mm}  
        & Domain walls
        & NG strings 
        & AH strings \\
        \noalign{\vskip 1mm}
        \hline
        \noalign{\vskip 1mm}
        $A_{\rm str}$ 
        & -- & $< 0.010$ & $< 0.026$ \\[1mm]
        $G\mu$ 
        & -- & $< 1.0 \times 10^{-7}$ & $< 1.6 \times 10^{-7}$ \\[1mm]
        \hdashline
        \noalign{\vskip 1.5mm}
        $A_{\rm DW}$ 
        & $< 4.3$ & -- & -- \\[1mm]
        $\sigma^{1/3}\;[\mathrm{MeV}]$
        & $< 1.3$ & -- & -- \\[1mm]
        \hline
        \hline
    \end{tabular}
    }
    \caption{95\% credible upper limits on $A_{\rm str}$ and $A_{\rm DW}$, together with the corresponding derived limits on the string and wall tensions, $G\mu$ and $\sigma^{1/3}$, for $L_f=1$. These bounds are obtained from \textit{Planck}$+$BK18 data, varying also the tensor-to-scalar ratio $r$.}
    \label{tab:constraints_Lf1}
\end{table}

\begin{figure}[!t]
    \centering
    \includegraphics[width=0.95\linewidth]{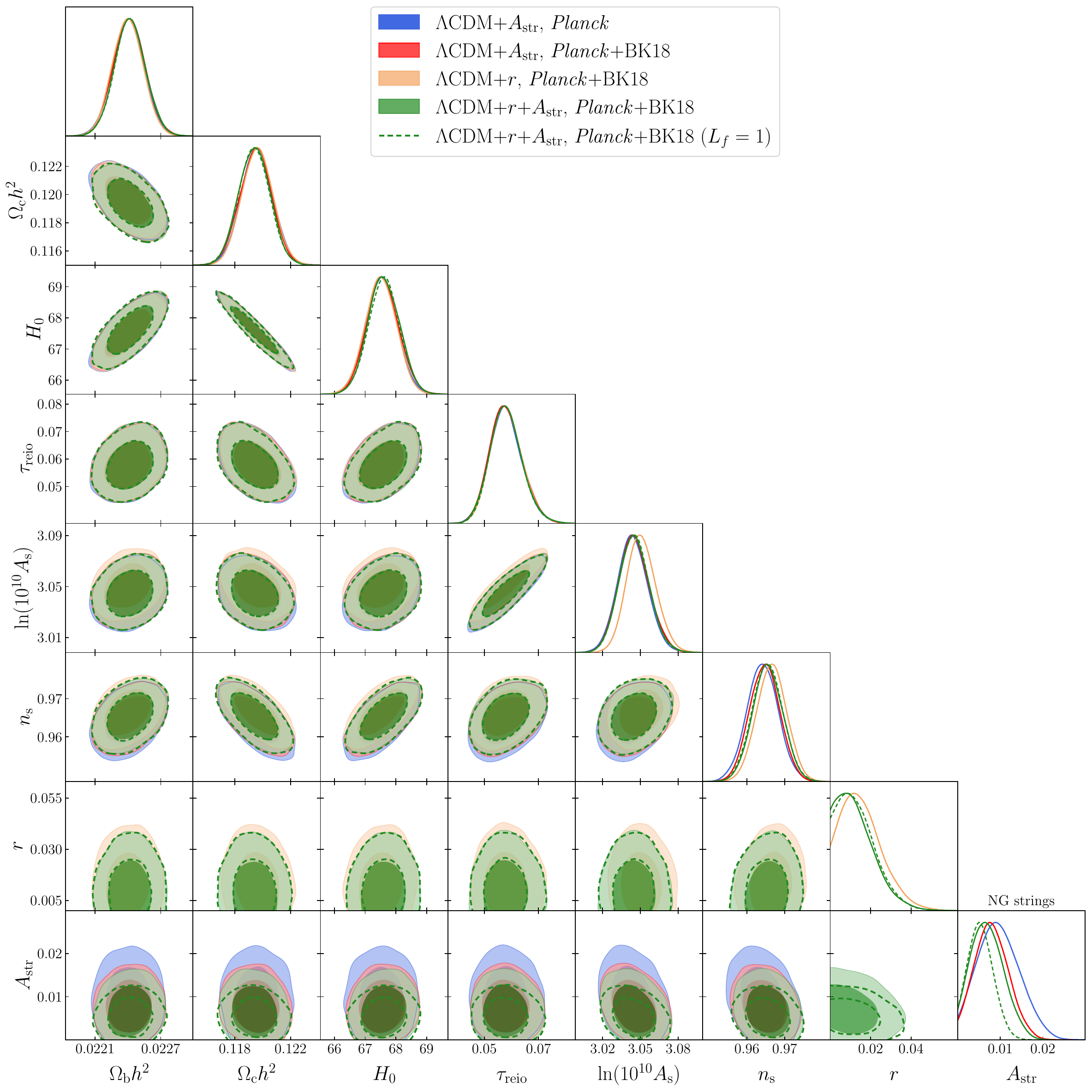}
    \caption{NG strings: full triangle plot showing one and two-dimensional posterior distributions for $A_{\rm str}$ and the cosmological parameters, for different models and dataset combinations.}
    \label{fig:Strings_NG_triangle_full}
\end{figure}

\begin{figure}[!t]
    \centering
    \includegraphics[width=0.95\linewidth]{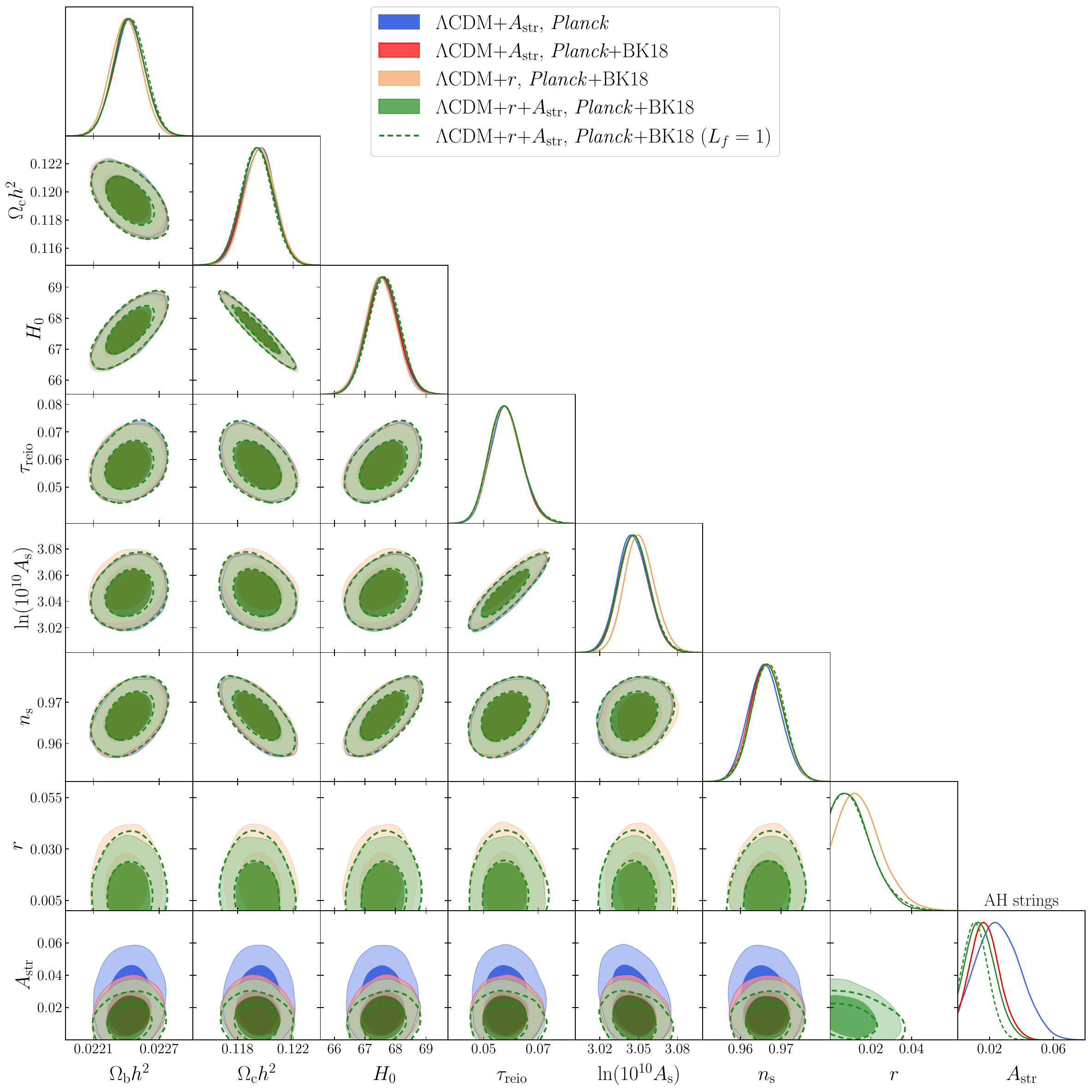}
    \caption{AH strings: full triangle plot showing one- and two-dimensional posterior distributions for $A_{\rm str}$ and the cosmological parameters, for different models and dataset combinations.}
    \label{fig:Strings_AH_triangle_full}
\end{figure}

\begin{figure}[!t]
    \centering
    \includegraphics[width=1\linewidth]{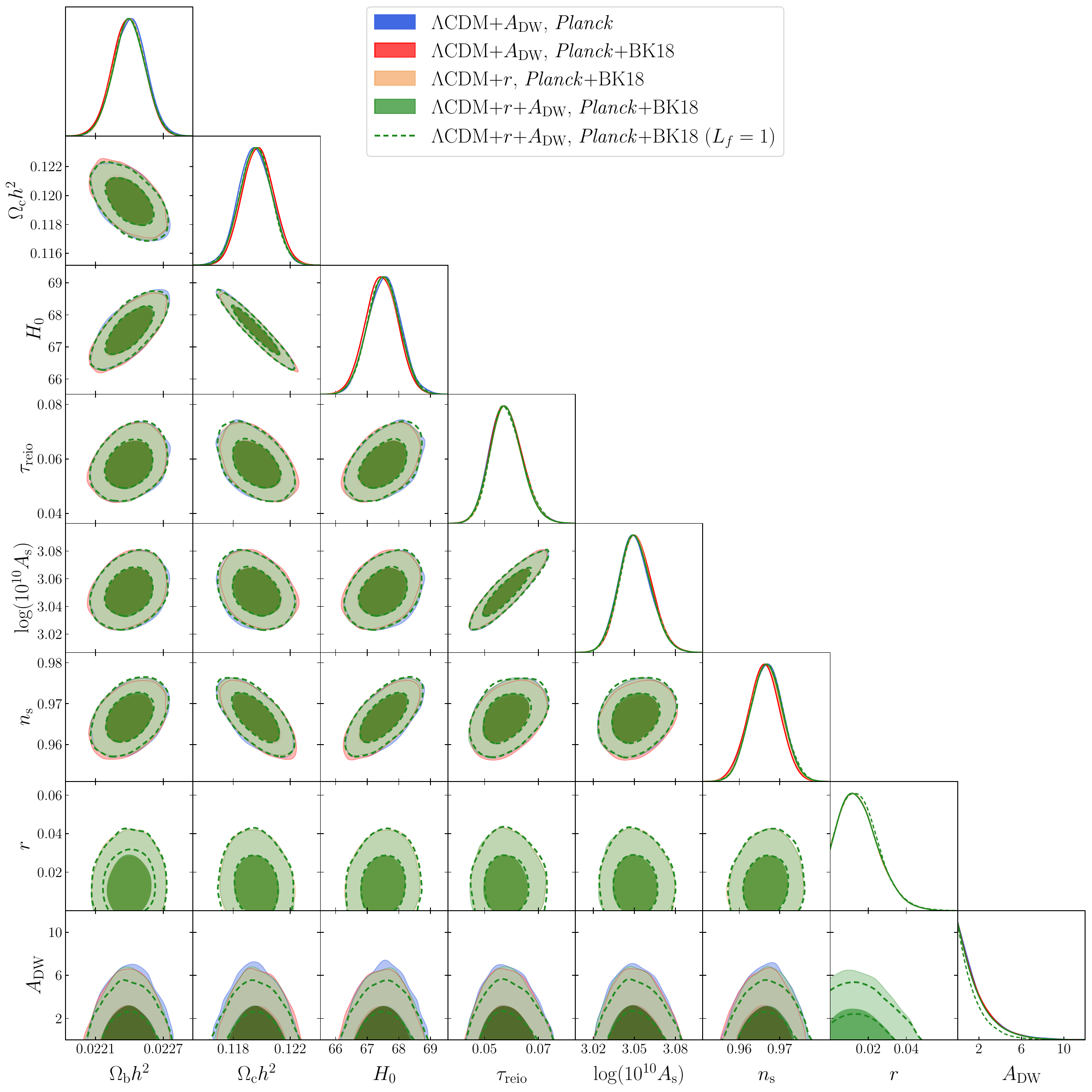}
    \caption{Domain walls: full triangle plot showing one and two-dimensional posterior distributions for $A_{\rm DW}$ and the cosmological parameters, for different models and dataset combinations.}
    \label{fig:DWs_stable_triangle}
\end{figure}

\clearpage

\section{Forecasted constraints for \textit{Planck}$+$SO and \textit{LiteBIRD}}
\label{app:SO}
\begin{figure}[!t]
    \centering
    \includegraphics[width=1\linewidth]{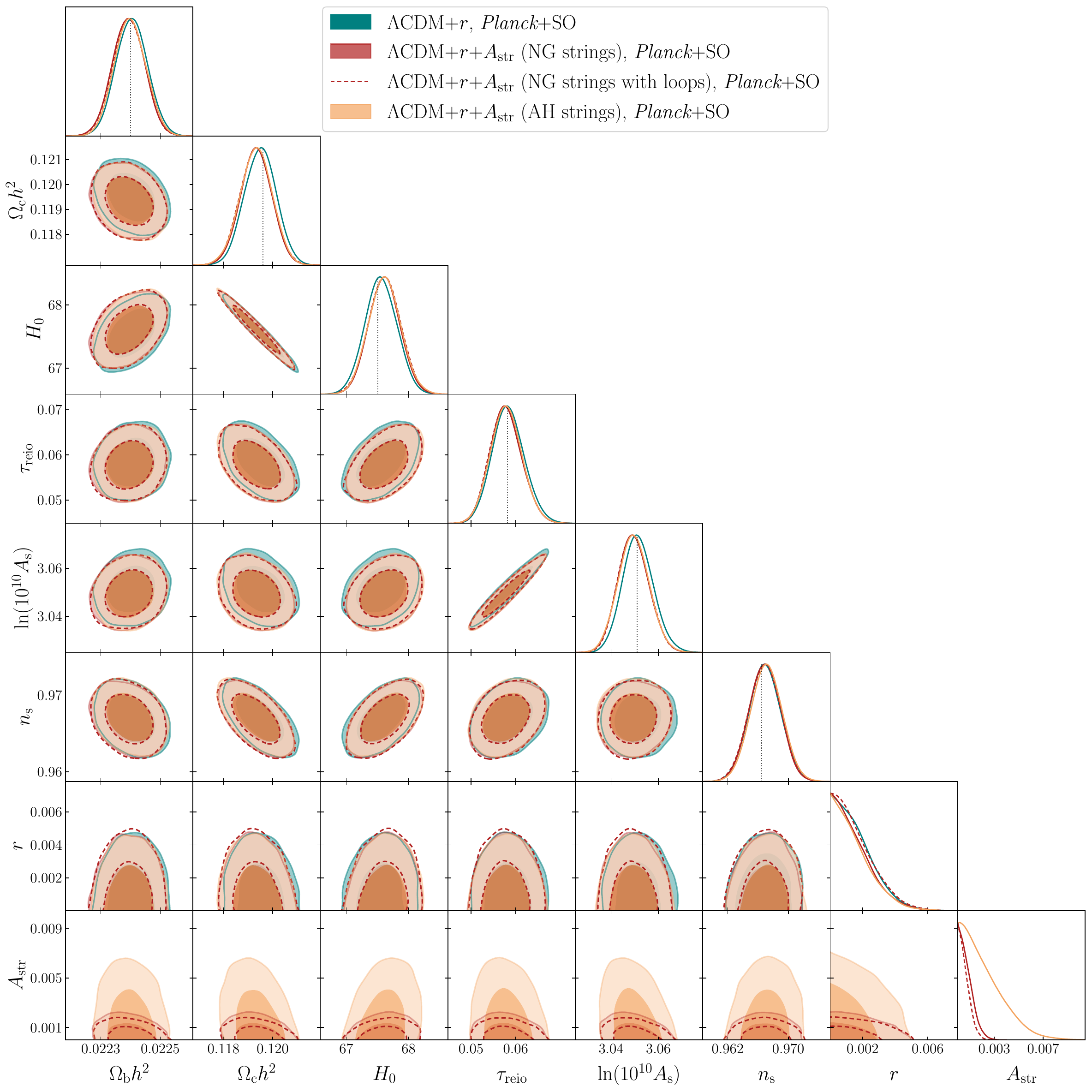}
    \caption{Full triangle plot showing one and two-dimensional posterior distributions for all parameters varied in our {\it Planck}$+$SO forecast analysis of cosmic string networks. Dashed vertical lines denote the fiducial $\Lambda$CDM parameter values used to generate the mock CMB spectra, and correspond to the mean values from our baseline {\it Planck}$+$BK18 analysis.}
    \label{fig:triangle_forecasts}
\end{figure}

\begin{figure}[!t]
    \centering
    \includegraphics[width=1\linewidth]{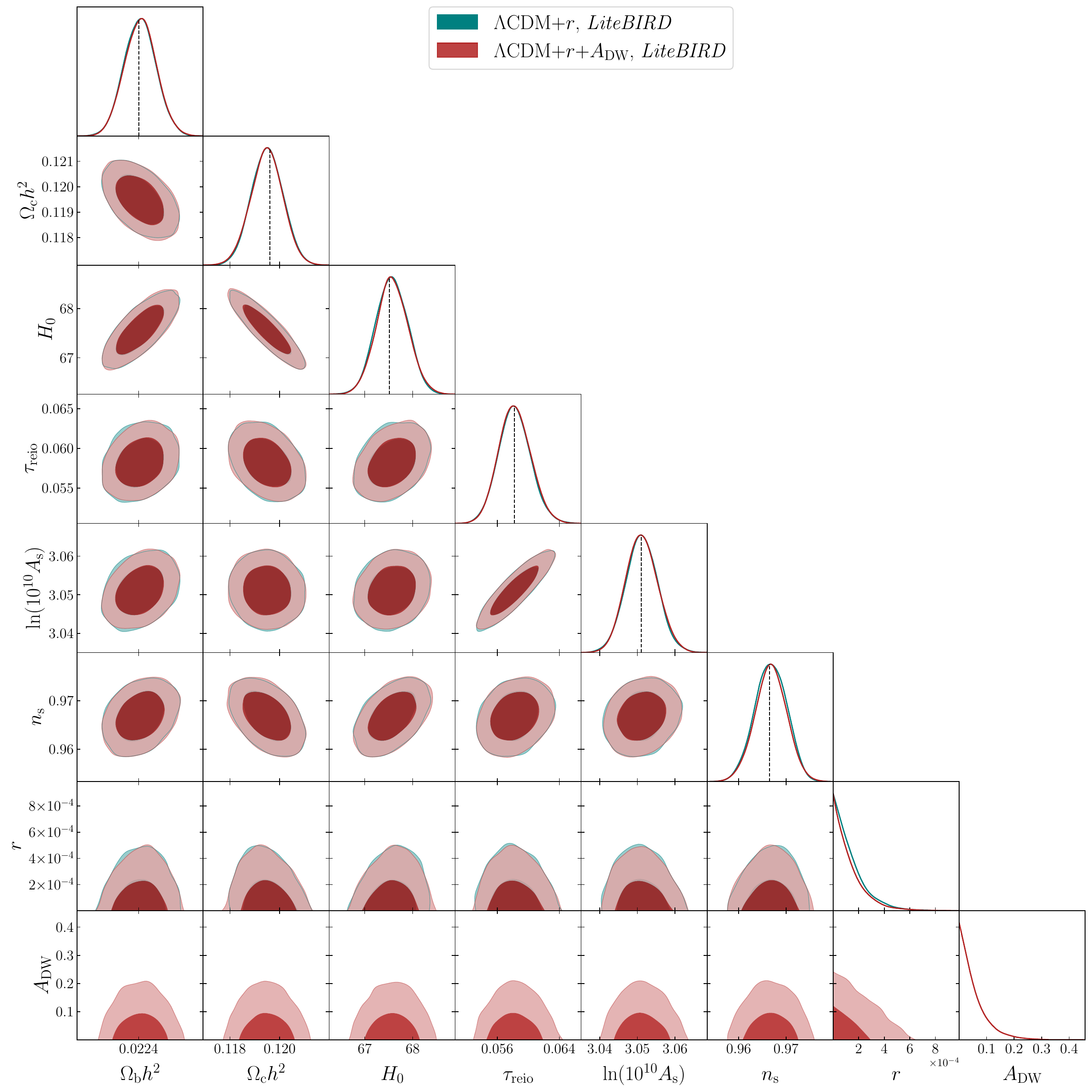}
    \caption{Full triangle plot showing one and two-dimensional posterior distributions for all parameters varied in our {\it LiteBIRD} forecast analysis of stable domain wall networks. Dashed vertical lines denote the fiducial $\Lambda$CDM parameter values used to generate the mock CMB spectra, as in figure~\ref{fig:triangle_forecasts}.}
    \label{fig:triangle_forecasts_LiteBIRD}
\end{figure}

This appendix presents the full triangle plots for the forecast analyses discussed in section~\ref{sec:forecasts}. 
Figure~\ref{fig:triangle_forecasts} shows the full triangle plot for the {\it Planck}$+$SO configuration, including the one- and two-dimensional posterior distributions for all cosmic string models considered in this work, namely Nambu-Goto strings with and without loops and Abelian-Higgs strings. For reference, we also show the forecasted constraints obtained for the baseline $\Lambda$CDM$+r$ model. The corresponding limits on all cosmological and string parameters are reported in table~\ref{tab:forecasts}.
Figure~\ref{fig:triangle_forecasts_LiteBIRD} displays the full triangle plot for the {\it LiteBIRD} forecast analysis of stable domain wall networks, including the $\Lambda$CDM$+r$ and $\Lambda$CDM$+r+A_{\rm DW}$ cases. The associated parameter constraints are summarized in table~\ref{tab:forecasts_Litebird}.

\begin{table}[!t]
    \centering
    \textbf{\small \textit{Planck}$+$SO} \\[0.6mm]
    \resizebox{\textwidth}{!}{
    \begin{tabular}{ c c c c c } 
        \hline
        \hline
        \noalign{\vskip 1mm}
        Parameter 
        & $\Lambda$CDM$+r$ 
        & $+$NG strings 
        & $+$NG strings with loops 
        & $+$AH strings \\
        \noalign{\vskip 1mm}
        \hline
        \noalign{\vskip 1mm}
        $\omega_b$ 
        & $0.022405 \pm 0.000053$ & $0.022393 \pm 0.000053$ & $0.022394 \pm 0.000053$ & $0.022399 \pm 0.000053$ \\[1mm]
        $\omega_c$ 
        & $0.11948 \pm 0.00063$ & $0.11933 \pm 0.00063$ & $0.11933 \pm 0.00065$ & $0.11933 \pm 0.00065$ \\[1mm]
        $H_0 \; [\mathrm{km}\,\mathrm{s}^{-1}\,\mathrm{Mpc}^{-1}]$ 
        & $67.56 \pm 0.25$ & $67.61 \pm 0.25$ & $67.61 \pm 0.26$ & $67.62 \pm 0.26$ \\[1mm]
        $\tau_{\rm reio}$ 
        & $0.0584^{+0.0033}_{-0.0037}$ & $0.0578^{+0.0033}_{-0.0037}$ & $0.0578^{+0.0033}_{-0.0038}$ & $0.0579 \pm 0.0035$ \\[1mm]
        $\ln(10^{10} A_s)$ 
        & $3.0515 \pm 0.0066$ & $3.0495 \pm 0.0065$ & $3.0496^{+0.0060}_{-0.0069}$ & $3.0496 \pm 0.0066$ \\[1mm]
        $n_s$ 
        & $0.9669 \pm 0.0021$ & $0.9669 \pm 0.0021$ & $0.9669 \pm 0.0021$ & $0.9671 \pm 0.0021$ \\[1mm]
        $r$ 
        & $< 3.8 \times 10^{-3}$ & $< 3.8 \times 10^{-3}$ & $< 3.9 \times 10^{-3}$ & $< 3.8 \times 10^{-3}$ \\[1mm]
        \hline
        \noalign{\vskip 1mm}
        $A_{\rm str}$ 
        & -- & $< 1.8 \times 10^{-3}$ & $< 1.4 \times 10^{-3}$ & $< 5.4 \times 10^{-3}$ \\[1mm]
        $G\mu$ 
        & -- & $< 4.2 \times 10^{-8}$ & $< 3.8 \times 10^{-8}$ & $< 7.3 \times 10^{-8}$ \\[1mm]
        \hline
        \hline
    \end{tabular}
    }
    \caption{Forecasted constraints on cosmic string networks for the {\it Planck}$+$SO configuration. $\Lambda$CDM parameters are quoted as 68\% credible intervals, while 95\% credible upper limits are reported for $r$ and $A_{\rm str}$ for the different string models considered.}
    \label{tab:forecasts}
\end{table}

\begin{table}[!t]
    \centering
    \textbf{\small \textit{LiteBIRD}} \\[0.75mm]
    \resizebox{0.6\textwidth}{!}{
    \begin{tabular}{ c c c } 
        \hline
        \hline
        \noalign{\vskip 1mm}
        Parameter & $\Lambda$CDM$+r$ & $+$Domain walls \\
        \noalign{\vskip 1mm}
        \hline
        \noalign{\vskip 1mm}
        $\omega_b$ & $0.02241 \pm 0.00012$ & $0.02241 \pm 0.00012$ \\[1mm]
        $\omega_c$ & $0.11950 \pm 0.00063$ & $0.11948 \pm 0.00064$ \\[1mm]
        $H_0 \; [\mathrm{km} \, \mathrm{s}^{-1} \, \mathrm{Mpc}^{-1}]$ & $67.56 \pm 0.33$ & $67.58 \pm 0.33$ \\[1mm]
        $\tau_{\rm reio}$ & $0.0582 \pm 0.0021$ & $0.0583 \pm 0.0021$ \\[1mm]
        $\ln(10^{10} A_s)$ & $3.0512 \pm 0.0043$ & $3.0512 \pm 0.0043$ \\[1mm]
        $n_s$ & $0.9668 \pm 0.0034$ & $0.9668 \pm 0.0034$ \\[1mm]
        $r$ & $< 0.40 \times 10^{-3}$ & $< 0.38 \times 10^{-3}$ \\[1mm]
        \hline
        \noalign{\vskip 1mm} 
        $A_{\rm DW}$ & -- & $< 0.16$ \\[1mm]
        $\sigma^{1/3} \; [\mathrm{MeV}]$ & -- & $< 0.74$ \\[1mm]
        \hline
        \hline
    \end{tabular}
    }
    \caption{Forecasted constraints on stable domain wall networks for {\it LiteBIRD}. $\Lambda$CDM parameters are quoted as 68\% credible intervals, while 95\% credible upper limits are reported for $r$ and for the domain wall tension.}
    \label{tab:forecasts_Litebird}
\end{table}

\clearpage

\bibliographystyle{JHEP}
\bibliography{bibliography.bib}

\end{document}